\documentclass[aps,prd,nofootinbib,amssymb,eqsecnum,notitlepage]{revtex4-1}
\usepackage{graphicx}
\usepackage{bm}
\usepackage{amsmath,amsthm,amssymb}
\usepackage{color}

\usepackage{amsfonts}
\usepackage{hyperref}
\usepackage{comment}

\begin{document}
\newcommand{\newc}{\newcommand}

\newc{\be}{\begin{equation}}
\newc{\ee}{\end{equation}}
\newc{\ba}{\begin{eqnarray}}
\newc{\ea}{\end{eqnarray}}
\newc{\bea}{\begin{eqnarray*}}
\newc{\eea}{\end{eqnarray*}}
\newc{\Mpl}{M_{\rm pl}}
\newc{\da}{\delta{A}}
\newc{\mm}[1]{\textcolor{red}{#1}}
\newcommand{\mmc}[1]{\textcolor{blue}{[MM:~#1]}}

\allowdisplaybreaks[1]

\title{Black hole perturbations in vector-tensor theories: The odd-mode analysis}

\author{
Ryotaro Kase$^{1}$,
Masato Minamitsuji$^{2}$, 
Shinji Tsujikawa$^{1}$, and 
Ying-li Zhang$^{1}$}

\affiliation{
$^1$Department of Physics, Faculty of Science, Tokyo University of Science, 1-3, Kagurazaka,
Shinjuku-ku, Tokyo 162-8601, Japan\\
$^2$Centro de Astrof\'{\i}sica e Gravita\c c\~ao  - CENTRA,
Departamento de F\'{\i}sica, Instituto Superior T\'ecnico - IST,
Universidade de Lisboa - UL, Av. Rovisco Pais 1, 1049-001 Lisboa, Portugal.
}

\date{\today}

\begin{abstract}
In generalized Proca theories with vector-field derivative couplings, a bunch of hairy black hole solutions have been
derived on a static and spherically symmetric background. 
In this paper, we formulate the odd-parity black hole perturbations in generalized Proca theories
by expanding the corresponding action up to second order 
and investigate whether or not
black holes with vector hair suffer 
ghost or Laplacian instabilities. We show that the models with cubic couplings $G_3(X)$, where  $X=-A_{\mu}A^{\mu}/2$ 
with a vector field $A_{\mu}$, do not provide any 
additional stability condition as in  
General Relativity.
On the other hand, the exact charged stealth Schwarzschild solution with a nonvanishing longitudinal vector component $A_1$, which originates from the coupling to the Einstein tensor $G^{\mu\nu}A_\mu A_\nu$
equivalent to the quartic coupling $G_4(X)$
containing a linear function of $X$, 
is unstable in the vicinity of the event horizon. 
The same instability problem also persists for hairy black holes arising from general quartic power-law couplings 
$G_4(X) \supset \beta_4 X^n$ with the nonvanishing $A_1$, 
while the other branch with $A_1=0$ can be consistent 
with conditions for the absence of ghost and 
Laplacian instabilities. We also discuss the case of other exact and numerical black hole solutions associated with intrinsic 
vector-field derivative couplings and show that 
there exists a wide range of parameter spaces in which 
the solutions suffer neither ghost nor Laplacian 
instabilities against odd-parity perturbations. 
\end{abstract}

\maketitle

\section{Introduction}

The observational discovery of late-time cosmic acceleration 
in 1998 \cite{SNIa} has led to the idea that extra degrees of freedom (DOFs) 
beyond those appearing in General Relativity (GR) 
and/or Standard Model of particle physics may be responsible for the acceleration \cite{review}. 
One of such extra DOFs is a scalar field 
minimally coupled to gravity--dubbed 
quintessence \cite{quin}.
If the scalar field has a direct coupling to the gravity sector, 
it is known that Horndeski theories \cite{Horndeski} are 
most general scalar-tensor theories with second-order equations of motion \cite{Horndeski11}.
Indeed, Horndeski theories have been applied to 
the late-time cosmic acceleration \cite{Horncosmo} 
as well as to the behavior of solutions 
in the Solar System \cite{Hornlocal}.

The scalar field is not the unique possibility for realizing the 
cosmic acceleration, but a vector field $A_{\mu}$
can also play a similar role. 
If we respect the $U(1)$-gauge symmetry of a 
vector field, 
however, there are no nontrivial derivative interactions with gravity responsible for
the cosmic acceleration \cite{Mukoh}. 
This situation is substantially modified 
in the presence of a vector field without the $U(1)$-gauge symmetry, i.e., a generalized Proca field. 
The action of generalized Proca theories with derivative 
couplings to gravity was derived in 
Refs.~\cite{Heisenberg,Tasinato,Allys,Jimenez2016} from the 
demand of keeping the equations of motion up to second 
order to avoid an Ostrogradski instability. 
Such new derivative interactions can give rise to the 
late-time cosmic acceleration \cite{DeFelicecosmo,obsig1}, 
while recovering the behavior
in GR around local objects with weak 
gravitational backgrounds \cite{DeFeliceVain,Naka17}.

In modified gravity theories, the deviation from GR can 
manifest itself in strong gravitational backgrounds like  
black holes (BHs) and neutron stars (NSs) \cite{Vitor,Yagi}.
The recent direct detections of gravitational waves (GWs) 
by Advanced LIGO/Virgo from binary BH mergers \cite{LIGO} and a binary NS merger 
together with its optical counterpart \cite{LIGO2}
began to offer the possibility for 
probing GR and its possible deviation in the nonlinear regime 
of gravity. 
In the Einstein-Maxwell system of GR, there is a no-hair conjecture
stating that the asymptotically flat and stationary BH solutions are described 
only by three parameters, i.e., mass, electric charge, and angular 
momentum \cite{Israel,Carter,Wheeler,Hawking}. 
The no-hair  property is valid for a canonical scalar field 
minimally coupled to gravity \cite{Chase,BekenPRL}
and also for a scalar field which has a direct 
coupling to the Ricci scalar \cite{Hawking72,Beken95,Soti12}.
Bekenstein \cite{Beken72} showed that,  
for a massive vector field theory with the broken $U(1)$ 
gauge symmetry, the vector field  must vanish due to the regularity on the horizon,
and hence the resulting solution is given by the stationary BHs in GR without the vector hair. 

In Horndeski theories, there exist BH solutions with 
scalar hair on the static and spherically symmetric background
for a scalar field $\phi$ 
coupled to a Gauss-Bonnet term \cite{Soti1,Soti2,GB1,GB2,GB3} 
or for a time-dependent scalar
$\phi=qt+\psi(r)$ with nonminimal derivative couplings to the Einstein 
tensor \cite{Babi16,Rinaldi:2012vy,Anabalon:2013oea,Minamitsuji:2013ura} 
(see also Refs.~\cite{Hui,Babi14,Koba14,Lefteris}). 
The latter corresponds to a stealth Schwarzschild BH 
solution with 
a nontrivial scalar hair, which was shown to be plagued 
by an instability problem against odd-parity
perturbations in the vicinity of the BH event 
horizon \cite{Kobayashi3}.

In the Einstein-Proca theory, 
the static and spherically symmetric 
BH solution corresponds to the Schwarzschild spacetime 
without vector hair \cite{Beken72}. 
On the other hand, in generalized Proca theories with derivative self-interactions and nonminimal derivative couplings to gravity, there exist a bunch of 
exact and numerical BH solutions 
with vector hair \cite{Tasinato1,Tasinato2,Minamitsuji,GPBH,GPBH2,Fan,Cisterna,Babichev17,Fan2}. 
The cubic and quartic derivative interactions, 
which include the vector Galileon as specific cases, 
give rise to BH solutions with a primary Proca hair \cite{GPBH,GPBH2}. 
For quintic power-law couplings the BH configuration 
regular throughout the horizon exterior does not exist, but 
there are hairy solutions with secondary Proca hair in the presence of 
sixth-order and intrinsic vector-mode couplings. 
Unlike Horndeski theories, the existence of temporal vector component besides longitudinal mode allows one to realize many hairy BH solutions without tuning the models. 
We also note that the large temporal vector component 
with cubic and quartic derivative couplings allows 
the possibility for realizing the mass of NSs larger than 
that in GR \cite{Tasinato2,KMT17}.

In this paper, we will formulate the odd-parity BH perturbations by extending the Regge-Wheeler formalism \cite{Regge:1957td,Zerilli:1970se}
and investigate whether or not BH solutions derived in 
Refs.~\cite{Tasinato1,GPBH,GPBH2} suffer ghost or Laplacian instabilities.
On the static and spherically symmetric background, the perturbations can be decomposed into
odd- and even-parity modes. 
In Horndeski theories, the stabilities of BH solutions against 
odd- and even-parity perturbations were investigated 
in Refs.~\cite{Kobayashi1,Kobayashi2,Kobayashi3} 
(see also Refs.~\cite{GB,Motohashi,EFT,GGGP}).
The analysis of odd-parity modes shows that the exact 
stealth BH solution with a time-dependent scalar is plagued 
by instabilities in the vicinity of the event horizon \cite{Kobayashi3}. 
In the present work, we generally derive 
conditions for the absence of ghost and 
Laplacian instabilities associated with
odd-parity perturbations 
in generalized Proca theories.

We will show that the charged stealth BH solution 
with a nonvanishing longitudinal component 
$A_1$, found 
for the coupling to the Einstein tensor 
$G^{\mu\nu}A_\mu A_\nu$ \cite{Tasinato1}
equivalent to the quartic coupling 
$G_4(X)$ containing a linear function of $X$,
does not simultaneously satisfy conditions for the absence of ghost and Laplacian instabilities in the vicinity of 
the event horizon. 
Such an instability problem against odd-parity perturbations 
also persists for the other general quartic power-law couplings $G_4(X)\supset\beta_4 X^n$ with $A_1 \neq 0$. 
For the other branch with $A_1=0$, the BH solutions 
arising from quartic power-law interactions can be 
consistent with the conditions for the absence of ghosts and Laplacian instabilities throughout the horizon exterior.

We will also investigate whether ghost and
Laplacian instabilities against odd-parity perturbations are present or not for hairy BHs arising from cubic, quintic, and intrinsic vector-mode couplings.
The cubic derivative interactions do not give rise to 
any additional stability conditions as in GR.
The quintic power-law couplings do not lead to 
background solutions regular throughout the horizon exterior, but the exact BH solution found in Ref.~\cite{GPBH2} 
suffer neither ghost nor Laplacian instabilities.
For the BH solutions associated with intrinsic vector-mode 
couplings, we will also show the existence of theoretically consistent parameter space.

Our paper is organized as follows.
In Sec.~\ref{sec2}, we show the equations
of motion in generalized Proca theories on the 
static and spherically symmetric background.
In Sec.~\ref{sec3}, we derive the second-order action 
of odd-parity perturbations as well as conditions for 
the absence of ghost and Laplacian instabilities. 
In Sec.~\ref{sec4}, we study the stability of 
exact BH solutions against odd-parity perturbations, 
including the quartic coupling $G_4(X)\supset X/4$ 
as a special case. 
In Sec.~\ref{sec5}, we extend the analysis to general quartic 
power-law couplings $G_4(X)\supset\beta_4 X^n$ for 
the two branches characterized by 
$A_1 \neq 0$ and $A_1=0$.
In Secs.~\ref{sec6} and \ref{sec7}, we will discuss the 
cases of hairy BH solutions realized by intrinsic 
vector-mode power-law couplings.
Sec.~\ref{consec} is devoted to conclusions.

\section{Background equations}
\label{sec2}

The action of generalized Proca theories with 
a vector field $A_{\mu}$ is given by \cite{Heisenberg,Tasinato,Allys,Jimenez2016}
\be
S=\int d^{4}x \sqrt{-g} 
\left( F+\sum_{i=2}^{6} \mathcal{L}_{i} \right)\,,
\label{action}
\ee
with 
\ba
\mathcal{L}_{2}&=& G_{2}(X, F)\,,
\label{L2}
\\
\mathcal{L}_{3}&=& G_{3}(X) \nabla_{\mu} A^{\mu}\,,
\label{L3}
\\
\mathcal{L}_{4}&=& G_{4}(X) R 
+ G_{4,X}(X)\left[ (\nabla_{\mu} A^{\mu})^{2} 
-  \nabla_{\mu} A_{\nu} \nabla^{\nu} A^{\mu}\right] \,,\\
\mathcal{L}_{5}&=& G_{5}(X) G_{\mu\nu} \nabla^{\mu} A^{\nu} 
- \frac{1}{6} G_{5,X} (X) \left[ (\nabla_{\mu} A^{\mu})^{3} 
- 3 \nabla_{\mu} A^{\mu} \nabla_{\rho} A_{\sigma} \nabla^{\sigma} A^{\rho} 
+ 2 \nabla_{\rho} A_{\sigma} \nabla^{\nu} A^{\rho} \nabla^{\sigma} A_{\nu} \right]
\notag
\\
&&-g_{5}(X) \tilde{F}^{\alpha\mu}\tilde{F}^{\beta}_{~\mu} \nabla_{\alpha} A_{\beta}\,,
\\
\mathcal{L}_{6}&=& G_{6}(X) L^{\mu\nu\alpha\beta} \nabla_{\mu} A_{\nu} \nabla_{\alpha} A_{\beta}
+\frac{1}{2} G_{6,X}(X) \tilde{F}^{\alpha\beta} \tilde{F}^{\mu\nu} \nabla_{\alpha} 
A_{\mu} \nabla_{\beta} A_{\nu}\,,
\label{L6}
\ea
where $g_{\mu\nu}$ is the four-dimensional metric tensor, 
$g$ is the determinant of $g_{\mu\nu}$, and 
$\nabla_\mu$, $R$, $G_{\mu\nu}$ 
represent the covariant derivative, 
the Ricci scalar, the Einstein tensor
associated with $g_{\mu\nu}$, respectively, 
and 
\ba
&&
X=-\frac{1}{2}A_{\mu}A^{\mu}\,,\qquad
F_{\mu \nu}=\nabla_{\mu}A_{\nu}-\nabla_{\nu}A_{\mu}\,,\qquad 
F=-\frac{1}{4}F_{\mu \nu}F^{\mu \nu}\,,\qquad  \notag\\ 
&& 
\tilde{F}^{\mu\nu}=\frac{1}{2} \mathcal{E}^{\mu\nu\alpha\beta} F_{\alpha\beta}\,,\qquad
L^{\mu\nu\alpha\beta}=\frac{1}{4} \mathcal{E}^{\mu\nu\rho\sigma} \mathcal{E}^{\alpha\beta\gamma\delta} R_{\rho\sigma\gamma\delta}\,.
\ea
Here, $\mathcal{E}^{\mu\nu\alpha\beta}$ is the Levi-Civita tensor satisfying the normalization  
$\mathcal{E}^{\mu\nu\alpha\beta}\mathcal{E}_{\mu\nu\alpha\beta}=-4!$, and $R_{\rho\sigma\gamma\delta}$ is the Riemann tensor. 
While the function $G_2$ is dependent on the two 
quantities $X$ and $F$, the functions $G_{3,4,5,6}$ and $g_5$ depend on $X$ alone with the notation of partial derivatives $G_{i,X} \equiv \partial G_{i}/\partial X$.

In Eq.~(\ref{L2}), we can also take into account the dependence of another term $Y=A^{\mu}A^{\nu}
{F_{\mu}}^{\alpha}F_{\nu \alpha}$ respecting the 
parity invariance \cite{obsig1}.
On the static and spherically symmetric spacetime
this quantity is expressed as $Y=4FX$ \cite{GPBH2}, 
so it is redundant to include the $Y$ dependence for 
discussing the background BH solutions. 
The dynamics of perturbations on the static and spherically symmetric background may be affected by the quantity $Y$, but we will not consider the explicit $Y$ dependence in 
$G_2$ throughout this paper.
The Lagrangians containing the functions $G_2=-2g_4(X)F$, 
$g_5(X)$, and $G_6(X)$ correspond to intrinsic 
vector-modes \cite{DeFelicecosmo,obsig1}. 
The generalized Proca theories given by the 
action (\ref{action}) lead to the second-order equations 
of motion with five propagating degrees of freedom.

We consider a static and spherically symmetric background 
given by the line element
\be
ds^{2} 
=\bar{g}_{\mu\nu}dx^\mu dx^\nu
=-f(r) dt^{2} +h^{-1}(r)dr^{2} + 
r^{2} \left( d\theta^{2}+\sin^{2}\theta\, 
d \varphi^{2} \right)\,,
\label{metric}
\ee
with the vector field in the form 
\be
\bar{A}_{\mu}=\left( A_0(r), A_1(r), 0, 0 \right)\,,
\label{vector_ansatz}
\ee
where $f$, $h$, $A_0$, and $A_1$ are functions of
the radial coordinate $r$. 
On the background (\ref{metric}), the $\theta$ and 
$\varphi$ components of spatial vector 
component $A_i$ need to vanish. 

In the following, we use the following notations
\be
X=X_0+X_1\,,\qquad 
X_0=\frac{A_0^2}{2f}\,,\qquad 
X_1=-\frac{hA_1^2}{2}\,.
\label{Xdef}
\ee
The gravitational equations of motion arising from 
the action (\ref{action}) are \cite{GPBH2}
\ba
\hspace{-1.cm}
& &
\left( c_{1} + \frac{c_{2}}{r} + \frac{c_{3}}{r^{2}} \right) h' 
+ c_{4} + \frac{c_{5}}{r} + \frac{c_{6}}{r^{2}}
=0\,, 
\label{be1} \\
\hspace{-1.cm}
& &-\frac{h}{f} \left( c_{1} + \frac{c_{2}}{r} + \frac{c_{3}}{r^{2}} \right) f' 
+ c_{7} + \frac{c_{8}}{r} + \frac{c_{9}}{r^{2}}
=0\,,
\label{be2}\\
\hspace{-1.cm}
& & \left( c_{10} + \frac{c_{11}}{r} \right) f'' + \left( c_{12} + \frac{c_{13}}{r} \right) f'^{2}
+ \left( \frac{c_{2}}{2 f} + \frac{c_{14}}{r} \right) f' h' + \left( c_{15} + \frac{c_{16}}{r} \right) f'
+\left( -\frac{c_{8}}{2 h} + \frac{c_{17}}{r} \right) h' + c_{18} + \frac{c_{19}}{r}=0\,,
\label{be3}
\ea
where the coefficients $c_{1,2,\cdots,19}$ are given in Appendix \ref{appa}, 
and a prime represents the derivative 
with respect to $r$. 
Varying the action (\ref{action}) with respect to 
$A_0$ and $A_1$, it follows that 
\ba
\hspace{-0.4cm}
& & 
rf \left[ 2fh(rA_0''+2A_0')+r(fh'-f'h)A_0' \right] (1+G_{2,F})
+r^2hA_0'^2 \left[ 2fhA_0''-(f'h-fh') A_0' \right] G_{2,FF}
-2 r^2f^2A_0 G_{2,X}
\notag\\
\hspace{-0.4cm}
& & 
-2 r^2fA_0' \left( fh^2A_1A_1' -hA_0A_0'+f'hX_0-fh'X_1\right) G_{2,XF}
-rfA_0 \left[ 2 rfhA_1'+(rf'h+rfh'+4fh)A_1 \right] G_{3,X}
\notag\\
\hspace{-0.4cm}
& & 
+4 f^2A_0 (rh'+h-1) G_{4,X}
-8 fA_0 \left[ rfh^2 A_1A_1'-(rf'h+rfh'+fh) X_1\right] G_{4,XX}
\notag\\
\hspace{-0.4cm}
& & 
-fA_0 \left[ f(3h-1)h'A_1+h(h-1) (f'A_1+2fA_1')   \right] G_{5,X}
-2 fhA_0X_1\left[2 fhA_1'+(f'h+fh')A_1 \right] G_{5,XX}
\notag\\
\hspace{-0.4cm}
& & 
-2 f \left[ f (3 h-1) h'A_0'+h (h-1) (2fA_0''-f'A_0') \right] G_6
-4 fh A_0'X_1 \left( h A_0 A_0'-2 fh^2 A_1 A_1'
-2 f'hX_0+2 fh'X_1 \right) G_{6,XX}
\notag\\
\hspace{-0.4cm}
& & 
-2 f \left[ 4 fh^2 X_1 A_0''-2 h (hX- X_0) f'A_0'
+2f (6h-1)h' X_1A_0'+h(h-1)A_0A_0'^2
-2 fh^2(3h-1) A_0'A_1A_1'\right] G_{6,X}
\notag\\
\hspace{-0.4cm}
& & 
-4 fh \left[ 2 rfh A_1 A_0''- \{(rf' h-3 rfh'-2fh) A_1-2 rfhA_1'\} A_0'  \right] g_5
\notag\\
\hspace{-0.4cm}
& & 
-4 rfh A_0' \left[ hA_0A_0'A_1+4 fhX_1A_1'-2 A_1(f'hX_0-fh'X_1)\right] g_{5,X}
=0\,,
\label{be4}
\ea
and 
\ba
\hspace{-0.4cm}
& & 
A_1 \left[ r^2fG_{2,X}-2 (rf'h+fh-f) G_{4,X}
+4h(rA_0 A_0'-rf' X-fX_1) G_{4,XX}
-hA_0'^2(3h-1) G_{6,X}-2h^2X_1 A_0'^2 G_{{6,{  XX}}}\right]
\nonumber \\
\hspace{-0.4cm}
& &=r [r(f' X-A_0 A_0')+4 fX_1] G_{3,X}
+2 f'hX_1G_{5,X}+(A_0 A_0'-f' X)\left[ (1-h)G_{5,X}-2 hX_1G_{5,XX}\right]
\nonumber \\
\hspace{-0.4cm}
& &
\hspace{.35cm}
-2rh A_0'^2( g_{5} +2 X_1 g_{5,X})\,.
\label{be5}
\ea
For the theories where only the functions $G_i(X)$ with 
even indices $i$ are present, the right hand side of Eq.~(\ref{be5}) 
vanishes. Then, there exists the branch satisfying $A_1=0$.
For the theories containing $G_i(X)$ with odd indices $i$, 
the general solution to Eq.~(\ref{be5}) corresponds to 
a nonvanishing longitudinal component. 
There are a bunch of exact and numerical BH solutions 
with $A_1 \neq 0$ and $A_1=0$ for the above system. 

\section{Second-order action for odd-parity perturbations}
\label{sec3}

On the static and spherically symmetric background (\ref{metric}), we consider small metric perturbations 
$h_{\mu \nu}$ with the vector-field perturbation 
$\delta A_{\mu}$. 
Then, the four-dimensional metric 
$g_{\mu \nu}$ and the vector field $A_{\mu}$ 
including perturbations are given, respectively, by 
\be
g_{\mu\nu}=\bar{g}_{\mu\nu}+h_{\mu\nu},
\qquad
A_\mu=\bar{A}_\mu+\delta A_\mu\,.
\ee
The perturbations on the background (\ref{metric}) can be 
decomposed into even- and odd-parity modes in the following 
manner. Under the rotation in two-dimensional sphere with 
the coordinates $\theta$ and $\varphi$, the metric perturbations 
$h_{tt}, h_{tr}, h_{rr}$ transform as scalars. 
Any scalar quantity $\Phi$ can be expressed in terms 
of the spherical harmonics $Y_{lm}(\theta, \varphi)$ as 
\be
\Phi(t,r,\theta, \varphi)=\sum_{l,m} 
\Phi_{lm}(t,r)Y_{lm}(\theta, \varphi)\,,
\label{Phidecom}
\ee
where $\Phi_{lm}$ is a function of $t$ and $r$. 
This scalar mode has the parity $(-)^l$ under the 
two-dimensional rotation.
The perturbations with parities $(-)^l$ and 
$(-)^{l+1}$ are called even-mode and odd-mode, 
respectively  \cite{Regge:1957td,Zerilli:1970se}. 
Under the two-dimensional rotation, the perturbations 
$h_{t\theta}, h_{t\varphi}, h_{r \theta}, 
h_{r \varphi}$ transform as 
vectors, while $h_{\theta \theta}, h_{\theta \varphi}, 
h_{\varphi \varphi}$ transform as tensors. 
We can decompose any vector field 
$V_a$ and any symmetric tensor 
$T_{ab}$ into the following forms \cite{Motohashi}:
\ba
V_a(t,r,\theta, \varphi) &=& \nabla_a \Phi_1
+E_{ab} \nabla^b \Phi_2\,,
\label{Va} \\
T_{ab}(t,r,\theta, \varphi) &=& \nabla_a \nabla_b \Psi_1+\gamma_{ab} 
\Psi_2+\frac{1}{2} \left( {E_a}^c \nabla_c \nabla_b \Psi_3
+{E_b}^c  \nabla_c \nabla_a \Psi_3 \right)\,,
\label{Tab} 
\ea
where the subscripts $a,b$ are either $\theta$ or $\varphi$, and
$\Phi_1, \Phi_2, \Psi_1, \Psi_2, \Psi_3$ are scalar 
quantities. The tensor $E_{ab}$ is given by 
$E_{ab}=\sqrt{\gamma}\, \varepsilon_{ab}$, where 
$\gamma$ is the determinant of two dimensional metric 
$\gamma_{ab}$ on the sphere and 
$\varepsilon_{ab}$ is the anti-symmetric symbol 
with $\varepsilon_{\theta\varphi}=1$. 
The terms containing the anti-symmetric tensor $E_{ab}$
in Eqs.~(\ref{Va}) and (\ref{Tab}) correspond to the odd-parity 
modes, whereas the other terms correspond to 
the even-parity modes. 
On using Eq.~(\ref{Phidecom}) in Eqs.~(\ref{Va}) and (\ref{Tab}), the vector $V_a$ and the tensor $T_{ab}$ 
can be expressed in terms of the sum of spherical harmonics 
and their $\theta, \varphi$ derivatives multiplied by functions 
containing the $(t,r)$ dependence over $l$ and $m$. 

In this paper, we focus on 
the odd-parity perturbations in generalized Proca theories.
{}From the above argument, the metric 
perturbations corresponding to the odd modes can be 
expressed in the forms
\ba
&&
h_{tt}=h_{tr}=h_{rr}=0\,,\label{htt} \\
&&
h_{ta}=\sum_{l,m}Q_{lm}(t,r)E_{ab}\partial^bY_{lm}(\theta,\varphi)\,,
\label{Qlm}\\
& &
h_{ra}=\sum_{l,m}W_{lm}(t,r)E_{ab}\partial^bY_{lm}(\theta,\varphi)\,,\\
&&
h_{ab}=\frac{1}{2}\sum_{lm}
U_{lm} (t,r)
\left[
E_{a}{}^c \nabla_c\nabla_b Y_{lm}(\theta,\varphi)
+ E_{b}{}^c \nabla_c\nabla_a Y_{lm}(\theta,\varphi)
\right]\,,
\label{Ulm}
\ea
where $Q_{lm}$, $W_{lm}$, $U_{lm}$ are functions 
of $t$ and $r$. 
The vector-field perturbation 
for the odd-parity sector is given by 
\ba
&&
\da_{t}=\da_{r}=0\,,\\
&&
\da_{a}=\sum_{l,m}\da_{lm}(t,r)E_{ab}\partial^bY_{lm}(\theta,\varphi)\,,
\label{vectorper}
\ea
where $\da_{lm}$ is a function of $t$ and $r$. 
All the perturbation variables are not necessarily physical 
because of the gauge degrees of freedom. 
Under the infinitesimal gauge transformation
$x_\mu\to x_\mu+\xi_\mu$ with 
\be
\xi_t=\xi_r=0,
\qquad
\xi_a=\sum_{lm}\Lambda_{lm}(t,r) E_{ab} 
\partial^b Y_{lm}(\theta,\varphi)\,,
\label{xivec}
\ee
the perturbations $Q_{lm}(t,r),W_{lm}(t,r), U_{lm}(t,r)$ in Eqs.~(\ref{Qlm})-(\ref{Ulm}) 
transform, respectively, as
\be
Q_{lm} \to Q_{lm}+\dot{\Lambda}_{lm}\,,\qquad
W_{lm} \to W_{lm}+\Lambda'_{lm}
-\frac{2}{r}\Lambda_{lm}\,,\qquad
U_{lm} \to U_{lm}+2\Lambda_{lm}\,,
\label{gaugetra}
\ee
where a dot represents the derivative with 
respect to $t$. 
For $l \geq 2$, we will choose the Regge-Wheeler gauge 
$U_{lm}=0$, after which we are left with two variables 
$Q_{lm}$ and $W_{lm}$.
For the dipole mode $l=1$, the perturbation $h_{ab}$ vanishes identically, so we need to handle it separately. 
We also note that the odd-parity perturbations do not exist
for the monopole mode $l=0$.

We substitute Eqs.~(\ref{htt})-(\ref{vectorper}) 
into the action (\ref{action}) and expand it up to 
second order in perturbations. 
In the actual computation, we confirm that it is sufficient to set 
$m=0$ and perform the integrals with respect to $\theta$ and 
$\varphi$. In Appendix \ref{appb}, we show two integrals of spherical harmonics used for our computation.
Performing the integrations with respect to $t$ and $r$ by parts, the resulting second-order action 
for odd-parity perturbations yields
\ba
S_{\rm odd}&=&\sum_{l,m} L \int dt dr\, 
{\cal L}_{\rm odd}\,, 
\label{oddact}
\ea
with
\be
L=l(l+1)\,,
\label{Ldef}
\ee
and
\ba
{\cal L}_{\rm odd}&=&r^2 \sqrt{\frac{f}{h}}
\biggl[
C_1\left(\dot{W}_{lm}-Q'_{lm}+\frac{2}{r}Q_{lm}\right)^2+2\left(C_2\dot{\da}_{lm}
+C_3\da'_{lm}+C_4\da_{lm} \right)
\left(\dot{W}_{lm}-Q'_{lm}+\frac{2}{r}Q_{lm}\right)
+C_5\dot{\da}_{lm}^2
\notag\\
&&
+C_6\dot{\da}_{lm}\da'_{lm}
+C_7{\da}_{lm}'^2+(L-2)\left( C_8W_{lm}^2+C_9 W_{lm}\da_{lm}+\frac{A_0}{f}C_9W_{lm}Q_{lm}
+C_{10}Q_{lm}^2
+C_{11}Q_{lm}\da_{lm} \right)
\notag\\
&&
+(LC_{12}+C_{13})\da_{lm}^2 \biggr]\,,
\label{oddLag}
\ea
where the background-dependent
coefficients $C_{1-13}$ are 
given in Appendix \ref{appc}. 

\subsection{The modes of $l \geq 2$}

For the modes of $l \geq 2$, there are two dynamical fields 
$W_{lm}$ and $\delta A_{lm}$.
The variable $Q_{lm}$ in Eq.~(\ref{oddLag}) is 
non-dynamical, so one can derive a constraint equation for it 
by varying the action (\ref{oddact}) with respect to $Q_{lm}$. 
Due to the presence of the term $Q_{lm}'^2$ in the action, however, the corresponding 
constraint equation cannot be explicitly solved for $Q_{lm}$. 
This problem can be tackled by using the method of a Lagrange multiplier (as already done in the context of 
scalar-tensor theories \cite{GB,Kobayashi1,Kobayashi2,EFT}).
We introduce a Lagrangian multiplier 
$\chi(t,r)$ and rewrite Eq.~(\ref{oddLag}) as 
\ba
\hspace{-0.8cm}
{\cal L}_{\rm odd}&=&r^2 \sqrt{\frac{f}{h}}
\left[ C_1\left\{2\chi\left(\dot{W}_{lm}-Q'_{lm}
+\frac{2}{r}Q_{lm}+\frac{C_2\dot{\da}_{lm}
+C_3\da'_{lm}+C_4\da_{lm}}{C_1}\right)-\chi^2\right\}
\right.
\notag\\
\hspace{-0.8cm}
&&\left.
-\frac{(C_2\dot{\da}_{lm}+C_3\da'_{lm}+C_4\da_{lm})^2}{C_1}+C_5\dot{\da}_{lm}^2+C_6\dot{\da}_{lm}\da'_{lm}+C_7{\da}_{lm}'^2
\right.
\notag\\
\hspace{-0.8cm}
&&\left.
+(L-2)\left(C_8W_{lm}^2+C_9W_{lm}\da_{lm}+\frac{A_0}{f}C_9W_{lm}Q_{lm}+C_{10}Q_{lm}^2+C_{11}Q_{lm}\da_{lm} \right)
+(LC_{12}+C_{13})\da_{lm}^2\right]. 
\label{LM}
\ea
Varying the corresponding action with respect to $\chi$
and substituting the equation of $\chi$ into Eq.~(\ref{LM}), 
one recovers the original Lagrangian (\ref{oddLag}). 
Variations of the Lagrangian (\ref{LM}) with respect to $W_{lm}$ and $Q_{lm}$ lead, 
respectively, to
\ba
&&
C_1\dot{\chi}-(L-2)\left[C_8W_{lm}
+\frac{C_9}{2}\left(\da_{lm}+\frac{A_0}{f}Q_{lm} 
\right)\right]=0\,,
\label{eqW}\\
&&
C_1\chi'+\frac{2rfhC_1'+(8fh+rf'h-rfh')C_1}{2rfh}\chi
+(L-2)\left(\frac{A_0C_9}{2f}W_{lm}+C_{10}Q_{lm}+\frac{C_{11}}{2}\da_{lm} \right)=0\,. 
\label{eqQ}
\ea
We solve Eqs.~(\ref{eqW}) and (\ref{eqQ}) 
for $W_{lm}$ and $Q_{lm}$, respectively, 
and then substitute these solutions 
into the Lagrangian (\ref{LM}). 
After integrating by parts, the Lagrangian \eqref{LM} 
can be written in the form  
\be
(L-2){\cal L}_{\rm odd}=r^2 \sqrt{\frac{f}{h}}\left( 
\dot{\vec{\mathcal{X}}}^{t}{\bm K}\dot{\vec{\mathcal{X}}}
+\dot{\vec{\mathcal{X}}}^{t}{\bm R}\vec{\mathcal{X}}'
+\vec{\mathcal{X}}'^{t}{\bm G}\vec{\mathcal{X}}'
+\dot{\vec{\mathcal{X}}}^{t}{\bm T}\vec{\mathcal{X}}
+\vec{\mathcal{X}}'^{t}{\bm S}\vec{\mathcal{X}}
+\vec{\mathcal{X}}^{t}{\bm M}\vec{\mathcal{X}}
\right)\,,
\label{LM2}
\ee
where $\vec{\mathcal{X}}^{t}=(\chi,\da_{lm})$, and 
${\bm {K,R,G,T,S,M}}$ are $2\times2$ matrices. 
Setting $\da_{lm}=0$, $A_0=q=\,$constant and $A_1=\psi'(r)$, Eq.~(\ref{LM2}) reduces to the second-order action 
for the odd-parity perturbations in shift-symmetric 
Horndeski theories with the linear time-dependent 
scalar field $\phi=qt+\psi(r)$.
We confirmed that, for the specific case $G_3=G_5=0$ 
in shift-symmetric Horndeski theories, 
the second-order Lagrangian derived above coincides 
with that derived in Ref.~\cite{Kobayashi3,Takahashi}.

The non-vanishing metric components of ${\bm {K,R,G}}$ 
are given by 
\ba
&&
K_{11}=q_1 \,,\qquad 
K_{22}=(L-2)q_2\,,
\label{kinetic}\\
&&
R_{11}=\frac{A_0C_9}{fC_{10}}K_{11}\,,\qquad 
R_{22}=\frac{(L-2)(C_1C_6-2C_2C_3)}{C_1}\,,\\
&&
G_{11}=\frac{C_8}{C_{10}}K_{11}\,,\qquad 
G_{22}=\frac{(L-2)(C_1C_7-C_3^2)}{C_1}\,,
\label{G1122}
\ea
where 
\ba
q_1&\equiv& \frac{4f^2C_1^2C_{10}}{A_0^2C_9^2-4f^2C_8C_{10}}\,,
\label{q1def}\\
q_2&\equiv& \frac{C_1C_5-C_2^2}{C_1}\,.
\label{q2def}
\ea
From Eq.~(\ref{kinetic}), the conditions for 
the absence of ghosts are
\ba
& & q_1>0\,,\label{NG1}\\
& & q_2>0\,.\label{NG2}
\ea

Assuming that the solutions of $\vec{\mathcal{X}}^{t}$ are 
in the form $e^{i(\omega t-kr)}$, the dispersion relation for large $\omega$ and $k$ yields
\be
{\rm det}(\omega^2{\bm K}-\omega k {\bm R}+k^2 {\bm G})=0\,.
\label{KRGre}
\ee
We derive the propagation speed $c_r=dr_*/d\tau$ along the radial direction in proper time outside the horizon 
($f>0$, $h>0$), where 
$d\tau=\sqrt{f}dt$ and $dr_*=dr/\sqrt{h}$.  
Since this is related to the propagation speed $\hat{c}_r$ 
in the coordinates $t$ and $r$ as $\hat{c}_r=\sqrt{fh}\,c_r$, 
we substitute the relation $\omega=\hat{c}_rk=\sqrt{fh}\,c_rk$
into Eq.~(\ref{KRGre}) and solve it for $c_r$. 
On using Eqs.~(\ref{kinetic})-(\ref{G1122}), 
this leads to the propagation speeds along 
the radial direction:
\ba
c_{r1}&=&\frac{A_0C_9\pm
\sqrt{{\cal F}_1}}{2f^{3/2}h^{1/2}C_{10}}\,,
\label{cr1}\\
c_{r2}&=&\frac{C_1C_6-2C_2C_3\pm
\sqrt{{\cal F}_2}}
{2f^{1/2}h^{1/2}C_1q_2}\,, 
\label{cr2}
\ea
where
\ba
{\cal F}_1 &\equiv& A_0^2C_9^2-4f^2C_8C_{10}\,,
\label{F1def}\\
{\cal F}_2 &\equiv& C_1^2(C_6^2-4C_7q_2)-4C_1C_3(C_2C_6-C_3C_5)\,.
\label{F2def}
\ea
Both $c_{r1}$ and $c_{r2}$ can be either positive or negative, 
depending on the direction along which the odd-parity 
perturbations propagate.
Since $c_{r2}$ does not arise for $\da_{lm}=0$, 
$c_{r1}$ and $c_{r2}$ correspond to the radial sound speeds 
arising from the gravity sector and the vector-field 
sector, respectively. 
The existence of real solutions to Eqs.~(\ref{cr1}) and (\ref{cr2}) 
requires the following two conditions:
\ba
&&{\cal F}_1\geq0\,,\label{NL1}\\
&&{\cal F}_2 \geq0\,,\label{NL2}
\ea
under which the small-scale Laplacian instability 
can be avoided.

In the large $\omega$ and $L$ limit, the two matrices 
${\bm M}$ and ${\bm T}$, besides ${\bm K}$,  
lead to contributions to the propagation 
speed $c_{\Omega}$ along the angular direction.
Picking up the dominant contributions to 
${\bm M}$ and ${\bm T}$ for the large $L$ limit, 
it follows that  
\be
M_{11}=-LC_1\,,\qquad 
M_{22}=L(L-2)D_1\,,\qquad
T_{12}=-T_{21}=-(L-2)D_2\,, 
\ee
where 
\be
D_1 \equiv C_{12}+\frac{fC_8C_{11}^2+C_9^2(fC_{10}-A_0C_{11})}{4fC_1^2C_{10}}q_1\,,\qquad 
D_2 \equiv C_2+\frac{C_9(2fC_{10}-A_0C_{11})}{4fC_1C_{10}}q_1\,.
\label{defD2}
\ee
The matrix component $M_{12}$ contains the term 
proportional to $L-2$, but it does not affect $c_{\Omega}$ 
derived below in the limit $L \to \infty$. 
On using the solution of $\vec{\mathcal{X}}^{t}$  
in the form $e^{i(\omega t-l \theta)}$, the dispersion 
relation is given by 
\be
{\rm det}(\omega^2{\bm K}-i\omega {\bm T}+{\bm M})=0\,.
\label{detKTM}
\ee
The propagation speed along the angular direction in 
proper time is $c_{\Omega}=r d\theta/d\tau=
\hat{c}_{\Omega}/\sqrt{f}$, where 
$\hat{c}_{\Omega}=rd\theta/dt$. 
We substitute the relation 
$\omega^2=\hat{c}_{\Omega}^2l^2/r^2
=c_{\Omega}^2 fl^2/r^2$ into Eq.~(\ref{detKTM}) 
and solve it for $c_{\Omega}^2$. 
Taking the $L \to \infty$ limit at the end, 
we obtain the two propagation speed squares as
\be
c_{\Omega \pm}^2=\frac{r^2}{2fq_1q_2}\left[C_1q_2-D_1q_1+D_2^2\pm
\sqrt{(C_1 q_2+D_1q_1)^2+D_2^2 
(2C_1 q_2-2D_1q_1+D_2^2)}\right]\,. 
\label{casq}
\ee
To avoid the Laplacian instability for large $L$,  
we require the conditions
\be
c_{\Omega \pm}^2\geq0\,. 
\label{casq2}
\ee
Thus, we have shown that the conditions (\ref{NG1}), (\ref{NG2}), 
(\ref{NL1}), (\ref{NL2}), and (\ref{casq2}) 
need to hold for avoiding ghost and Laplacian instabilities. 

\subsection{The dipole mode $l=1$}

The Regge-Wheeler gauge adopted for $l \geq 2$
corresponds to the gauge choice eliminating the perturbation $h_{ab}$. For the dipole mode $l=1$, however, 
the perturbation $h_{ab}$ vanishes identically, so the 
gauge is not fixed in Eq.~(\ref{oddact}). 
For $l=1$, i.e., $L=2$, the terms multiplied by the 
factor $(L-2)$ in the Lagrangian (\ref{oddLag}) vanish.

We recall that, under the gauge transformation 
$x_{\mu} \to x_{\mu}+\xi_{\mu}$ with Eq.~(\ref{xivec}), 
the perturbations $Q_{lm}$ and $W_{lm}$ transform as 
Eq.~(\ref{gaugetra}). 
For $l=1$, we choose the gauge in which $W_{lm}$ vanishes, 
such that 
\be
\Lambda_{1m} (t,r)=-r^2\int d\tilde{r}
\frac{W_{1m}(t,\tilde{r})}{\tilde{r}^2}
+r^2{\cal C}(t)\,, 
\label{gauge}
\ee
where ${\cal C}(t)$ is an arbitrary function of $t$ corresponding 
to a residual gauge degree of freedom. 
Varying the action (\ref{oddact}) with respect to $W_{1m}$ and $Q_{1m}$, 
and setting $W_{1m}=0$ at the end, it follows that 
\ba
\dot{\cal E} &=&0\,,\label{Eeq1}\\
\left(r^2{\cal E}\right)'&=&0\,,\label{Eeq2} 
\ea
where 
\be
{\cal E} \equiv r^2 \sqrt{\frac{f}{h}} 
\left[ C_1\left(Q'_{1m}-\frac{2}{r}Q_{1m}\right)-\left( C_2\dot{\da}_{1m}
+C_3\da'_{1m}+C_4\da_{1m}\right) \right]\,.
\label{calE}
\ee
The solutions to Eqs.~(\ref{Eeq1}) and (\ref{Eeq2}) 
are given by 
\be
{\cal E}=\frac{{\cal C}_1}{r^2}\,,
\label{calEso}
\ee
where ${\cal C}_1$ is an integration 
constant\footnote{Integrating Eq.~(\ref{calEso}) with 
Eq.~(\ref{calE}) to solve for $Q_{1m}$, it follows that 
\be
Q_{1m}=r^2\int d \tilde{r} \frac{1}{C_1 \tilde{r}^2} 
\left( \frac{{\cal C}_1}{\tilde{r}^4}\sqrt{\frac{h}{f}}
+C_2\dot{\da}_{1m}+C_3\da'_{1m}+C_4\da_{1m} \right)
+r^2{\cal C}_2(t) \,,
\notag
\ee
where ${\cal C}_2(t)$ is a time-dependent gauge mode. 
This gauge mode can be eliminated 
by setting the residual gauge degree of freedom 
$\Lambda_{1m}=r^2{\cal C}(t)$ to be 
${\cal C}(t)=\int dt\,{\cal C}_2(t)$.
In the case where ${\delta A}_{1m}=0$,
$Q_{1m}$ does not depend on  time,
and the integration constant ${\cal C}_1$
is related to the angular momentum of
a slowly rotating BH \cite{Kobayashi1,Kobayashi3}.
}. 
Since the $Q_{1m}$-dependent terms in the Lagrangian 
(\ref{oddLag}) appear only in the form of $Q_{1m}'-2Q_{1m}/r$, one can eliminate those terms 
by using the solution (\ref{calEso}) with Eq.~(\ref{calE}). 
After the integration by parts, the 
Lagrangian (\ref{oddLag}) reduces to 
\ba
{\cal L}_{\rm odd}=
&&
r^2 \sqrt{\frac{f}{h}} \Bigg[ 
\frac{C_1C_5-C_2^2}{C_1}\dot{\da}_{1m}^2
+\left(C_6-\frac{2C_2C_3}{C_1} \right)\dot{\da}_{1m}\da'_{1m} 
+\left(C_7-\frac{C_3^2}{C_1} \right)\da_{1m}'^2
\notag\\
&&
\hspace{1.5cm}
-\frac{2C_3C_4}{C_1}\da'_{1m}\da_{1m}
+\left(2C_{12}+C_{13}-\frac{C_4^2}{C_1}\right)\da_{1m}^2
+\frac{h\,{\cal C}_1^2}{C_1r^8f}
\Bigg]\,. 
\label{Ldipole2}
\ea
This result shows that only the vector perturbation 
$\da_{1m}$ propagates. The coefficient of 
$\dot{\da}_{1m}^2$ coincides with 
$q_2$ given by Eq.~(\ref{q2def}). 
Moreover, from the first three 
terms in Eq.~(\ref{Ldipole2}), one can show that the 
radial propagation speed of $\da_{1m}$ is the same 
as $c_{r2}$ given by Eq.~(\ref{cr2}). 
Thus, the dipole perturbation does not 
give rise to any additional condition for the absence of ghost and Laplacian instabilities
to those derived for 
$l\geq2$.

\subsection{GR and cubic couplings $G_3(X)$}\label{GRG3}

In GR with an electromagnetic field described by the Lagrangian $F$, 
the functions $G_i(X)$ in the action (\ref{action}) are 
\be
G_4=\frac{M_{\rm pl}^2}{2}\,,\qquad 
G_2=G_3=G_5=G_6=g_5=0\,,
\label{GRfun}
\ee
where $M_{\rm pl}$ is the reduced Planck mass. 
Integrating Eqs.~(\ref{be1})-(\ref{be5}) with respect to $r$ 
and using the boundary conditions $f=h=1$ at $r \to \infty$, 
we obtain the Reissner-Nordstr\"{o}m (RN)  solution:
\be
f=h=1-\frac{2M}{r}+\frac{Q^2}{2M_{\rm pl}^2r^2}\,,
\qquad A_0=P+\frac{Q}{r}\,,
\label{RNso}
\ee
with $A_1$ unfixed, where $M$ and $Q$ are 
mass and electric charge, respectively,
and $P$ is an arbitrary constant. 
The odd-parity perturbations about the RN solution were originally studied in Ref.~\cite{Moncrief}.
Substituting the functions (\ref{GRfun}) into Eqs.~(\ref{q1def})-(\ref{q2def}), 
(\ref{cr1})-(\ref{F2def}), and (\ref{casq}), it follows that 
\be
q_1=\frac{\Mpl^2h}{4f^2}\,,\qquad 
q_2=\frac{1}{2r^2f}\,,\qquad 
{\cal F}_1=\frac{M_{\rm pl}^4fh}{4r^8}\,,\qquad
{\cal F}_2=\frac{M_{\rm pl}^4h^3}{16r^8f^3}\,,\qquad
c_{r1}^2=c_{r2}^2=c_{\Omega+}^2=c_{\Omega-}^2=1\,,
\label{GR} 
\ee
where $f=h$.
Since all these quantities are positive outside the event horizon, 
the RN solution (\ref{RNso}) suffers neither ghost nor Laplacian instabilities. 

Let us then consider the cubic coupling $G_3(X)$ in the presence of 
the Einstein-Hilbert term, i.e.,
\be
G_3=G_3(X)\,,\qquad 
G_4=\frac{M_{\rm pl}^2}{2}\,,\qquad 
G_2=G_5=G_6=g_5=0\,. 
\label{GRfun2}
\ee
The $G_3$-dependent terms appear only via $C_{13}$ in Eq.~(\ref{oddLag}), 
which does not affect Eqs.~(\ref{q1def}), (\ref{q2def}), (\ref{cr1}), (\ref{cr2}), 
and (\ref{casq}). 
Hence the quantities $q_1,q_2,c_{r1},c_{r2},c_{\Omega}^2$ are not modified 
relative to those in GR given by Eq.~(\ref{GR}). Thus, the hairy BH solutions  
arising from the cubic couplings \cite{GPBH,GPBH2} satisfy the conditions for the absence of ghost and Laplacian instabilities against 
odd-parity perturbations. 

\section{Exact solutions}
\label{sec4}

In this section, we study whether or not the exact BH solutions 
obtained in Refs.~\cite{Tasinato1,GPBH,GPBH2} are 
free from ghost and Laplacian instabilities against 
odd-parity perturbations. 
These exact solutions were derived
by imposing the following two relations:
\be
f=h\,,\qquad X=X_c\,, 
\label{excon}
\ee
where $X_c$ is a constant. On using Eq.~(\ref{Xdef}), 
the latter condition of Eq.~(\ref{excon}) translates to 
\be
A_1=\epsilon \frac{\sqrt{A_0^2-2fX_c}}{f}\,,
\label{A1con}
\ee
where $\epsilon= \pm 1$.

The exact solution found in Ref.~\cite{Tasinato1} 
follows from the coupling 
$G_4(X)=\Mpl^2/2+X/4$, $G_2=G_3=G_5=G_6=g_5=0$.  
Imposing the two conditions (\ref{excon}), 
a family of exact BH solutions was obtained in 
Refs.~\cite{GPBH,GPBH2} in the presence 
of derivative couplings other than $G_4(X)$.
In the following, we first study the stability of 
the exact BH solution present for the quartic coupling, 
and then proceed to the cases of other exact solutions.

\subsection{Quartic coupling}
\label{exG4}

For general quartic couplings $G_4(X)$, 
Eq.~(\ref{be5}) gives 
\be
A_1 \left[ f \{ hrf'+ (h-1)f\}G_{4,X}
-h (A_1^2 ff'h r+A_1^2 f^2 h-A_0^2 f'r
+2A_0A_0' fr)G_{4,XX} \right]=0\,.
\label{be5exG4}
\ee
There are two branches characterized 
by $A_1 \neq 0$ and $A_1=0$. 
The exact solution derived in Ref.~\cite{Tasinato1} 
belongs to the former one, so we first study this case and then proceed to the latter branch.

\subsubsection{$A_1\neq0$}
\label{exG4A1neq0}

If the second derivative $G_{4,XX}$ obeys the relation  
$G_{4,XX}(X_c)=0$ with (\ref{excon}), then  
Eq.~(\ref{be5exG4}) reduces to $rf'+f-1=0$. 
This is integrated to give $f=h=1-2M/r$, where 
$M~(>0)$ is an integration constant.
There exists an exact solution consistent with 
Eqs.~(\ref{be1})-(\ref{be4}) for $G_{4,X}(X_c)=1/4$.
Then, for the quartic coupling given by 
\be
G_4(X)=G_4(X_c)+\frac{1}{4} (X-X_c) 
+\sum_{n=3} b_n(X-X_c)^n\,, 
\label{func1exG4}
\ee
we have the exact solution
 \cite{Tasinato1,GPBH,GPBH2}
\be
f=h=1-\frac{2M}{r}\,,\qquad 
A_0=P+\frac{Q}{r}\,,\qquad 
A_1=\epsilon \frac{\sqrt{2P(MP+Q)r+Q^2}}{r-2M}\,, 
\label{sol1exG4}
\ee
where $P$ and $Q$ are integration 
constants with $X_c=P^2/2$.
In the following, we choose $\epsilon=+1$ 
without loss of generality. 
In spite of the existence of charge $Q$ and nonvanishing longitudinal vector component $A_1$, the metric 
components are the same as those of the 
Schwarzschild metric. We call this the
charged stealth Schwarzschild solution. 
The model $G_4(X)=\Mpl^2/2+X/4$, 
which is equivalent to the Lagrangian 
${\cal L}=M_{\rm pl}^2R/2 +F+(1/4)G^{\mu\nu}
A_\mu A_\nu$ studied in Ref.~\cite{Tasinato1}, 
corresponds to the special case of 
Eq.~(\ref{func1exG4}) with
the functions $G_4(X_c)=M_{\rm pl}^2/2+X_c/4$ 
and $b_n=0$ for $n \geq 3$.

Let us study whether the solution (\ref{sol1exG4}) satisfies 
the stability conditions derived 
in Sec.~\ref{sec3} outside the 
event horizon ($r>2M$).
Due to the complexities of such conditions 
for general $r$, we focus on those in the vicinity 
of the horizon.
Substituting Eqs.~(\ref{func1exG4})-(\ref{sol1exG4}) 
into Eqs.~(\ref{q1def})-(\ref{q2def}) and then expanding them around $r=2M$, the quantities $q_1$ and $q_2$
reduce, respectively, to 
\ba
q_1&=& \frac{[P^2-4G_4(X_c)]
(2MP+Q)^2}{32G_4(X_c)(r-2M)^2}
+{\cal O}((r-2M)^{-1})\,,
\label{q1e}\\
q_2&=& \frac{(2MP+Q)^2}
{32M^2 [P^2-4G_4(X_c)](r-2M)^2}
+{\cal O}((r-2M)^{-1})\,. 
\label{q2e}
\ea
The two no-ghost conditions (\ref{NG1}) and (\ref{NG2}) 
are satisfied for 
\be
P^2>4G_4(X_c) \quad {\rm and} \quad G_4(X_c)>0\,.
\label{noghostTa}
\ee

The two quantities (\ref{NL1}) and (\ref{NL2}) 
associated with the radial propagation speeds are given by 
\ba
{\cal F}_1 &=& -\frac{[P^2-4G_4(X_c)]G_4(X_c)}{4096M^{10}}(r-2M)^2+
{\cal O}((r-2M)^{3})\,,\\
{\cal F}_2 &=& \frac{[P^2-4G_4(X_c)]
[3P^2-16G_4(X_c)]}{65536M^8}
+{\cal O}(r-2M)\,. 
\label{q1q2exG4}
\ea
Under the conditions (\ref{noghostTa}) for 
the absence of ghosts, we have ${\cal F}_1<0$
and hence the propagation speed $c_{r1}$ is imaginary. 
The other condition ${\cal F}_2 \geq 0$ is satisfied 
for $P^2 \geq 16G_4(X_c)/3$.

For general quartic couplings $G_4(X)$,
the term $D_2$ defined in Eq.~(\ref{defD2}) identically vanishes, so the two branches of Eq.~(\ref{casq}) reduce to 
$c_{\Omega+}^2={C_1r^2}/(fq_1)$ and $c_{\Omega-}^2=-{D_1r^2}/(fq_2)$, respectively. 
Around the event horizon, it follows that 
\ba
c_{\Omega+}^2
&=& -\frac{8MG_4(X_c)}{(2MP+Q)^2}(r-2M)
+{\cal O}((r-2M)^{2})\,,\label{cOG4}\\
c_{\Omega-}^2
&=& \frac{2M[3P^2-16G_4(X_c)]}
{(2MP+Q)^2}(r-2M)+{\cal O}((r-2M)^{2})\,.
\ea
While $c_{\Omega-}^2$ is positive for $P^2 \geq 16G_4(X_c)/3$, 
the other propagation speed squared $c_{\Omega+}^2$
is always negative under 
the latter condition of Eq.~(\ref{noghostTa}). 
In other words, from Eqs.~(\ref{q1e}), 
(\ref{q2e}), and (\ref{cOG4}), we obtain
\be
q_1 q_2 c_{\Omega+}^2=
-\frac{(2MP+Q)^2}{128M(r-2M)^3}
+{\cal O}((r-2M)^{-2})\,,
\label{qcpro}
\ee
which is negative.
Thus, under the no-ghost conditions $q_1>0$ and $q_2>0$, 
the exact BH solution (\ref{sol1exG4}) 
present for the quartic coupling (\ref{func1exG4})
is plagued by the Laplacian instability $c_{\Omega+}^2<0$ and 
by the problem of imaginary $c_{r1}$ around the 
event horizon.

\subsubsection{$A_1=0$}
\label{exG4A10}

We proceed to the other branch characterized by $A_1=0$. 
Under the two conditions (\ref{excon}), 
we can exactly solve Eqs.~(\ref{be1})-(\ref{be4}) for  
$G_{4,X}(X_c)=0$ and $G_{4}(X_c)=X_c/2$.
Thus, for the quartic coupling given by 
\be
G_4(X)=\frac{X_c}{2}
+\sum_{n=2} b_n\left( X-X_c \right)^n\,,
\label{func2exG4}
\ee
there exists the following extremal RN BH 
solution \cite{GPBH,GPBH2}:
\be
f=h=\left(1-\frac{M}{r}\right)^2\,,\qquad 
A_0=P\left(1-\frac{M}{r}\right)\,,\qquad 
A_1=0\,,
\label{sol2exG4}
\ee
where $P^2=2X_c$.
Substituting Eqs.~(\ref{func2exG4})-(\ref{sol2exG4}) into Eqs.~(\ref{q1def})-(\ref{q2def}), 
(\ref{cr1})-(\ref{F2def}), and (\ref{casq}), 
we obtain
\be
q_1=\frac{P^2}{8f}\,,\qquad 
q_2=\frac{1}{2r^2f}\,,\qquad 
{\cal F}_1=\frac{P^4f^2}{16r^8}\,,\qquad
{\cal F}_2=\frac{P^4}{64r^8}\,,\qquad
c_{r1}^2=c_{r2}^2=c_{\Omega}^2=1\,.
\ee
Since all these quantities are positive, the exact BH solution (\ref{sol2exG4}) suffers neither ghost nor Laplacian instabilities.

\subsection{Quintic coupling}
\label{exG5}

For the quintic coupling 
\be
G_5(X)=G_5(X_c)+\sum_{n=2} b_n 
\left( X-X_c \right)^n
\label{funcexG5}
\ee
with $X_c=M_{\rm pl}^2$, the following 
exact RN solution with charge $Q$
is present \cite{GPBH,GPBH2}:
\be
f=h=1-\frac{2M}{r}+\frac{Q^2}{2M_{\rm pl}^2r^2}\,,\qquad
A_0=-\frac{2MM_{\rm pl}^2}{Q}+\frac{Q}{r}\,,\qquad
A_1=\epsilon \frac{2M_{\rm pl}^3 
\sqrt{2(2M^2M_{\rm pl}^2-Q^2)}\,r^2}
{Q[2M_{\rm pl}^2r(2M-r)-Q^2]}\,.
\label{solexG5}
\ee
This has two branches: (i) $A_1 \neq 0$ 
(realized for $2M^2M_{\rm pl}^2>Q^2$), and 
(ii) $A_1=0$ (realized for $2M^2M_{\rm pl}^2=Q^2$).

Substituting Eqs.~(\ref{funcexG5})-(\ref{solexG5}) 
into Eqs.~(\ref{q1def})-(\ref{q2def}), 
(\ref{cr1})-(\ref{F2def}), and (\ref{casq}), 
we find that the quantities 
$q_1,q_2,{\cal F}_1, {\cal F}_2, c_{r1}^2,c_{r2}^2,
c_{\Omega \pm}^2$ are the same as those 
in GR given by Eq.~(\ref{GR}). 
Hence the exact solution 
(\ref{solexG5}) is plagued by neither ghost nor Laplacian instabilities.

\subsection{Sixth-order coupling}
\label{exG6}

For general sixth-order couplings $G_6(X)$, 
Eq.~(\ref{be5}) reduces to 
\be
A_1 A_0'^2 \left[ A_1^2h^2 G_{6,XX}
+\left(1-3h\right)G_{6,X}  \right]=0\,.
\label{be5exG6}
\ee
There are two non-trivial branches characterized 
by (i) $A_1=0$ and (ii) $A_0'=0$. 
In the following, we study the stability of 
solutions in each branch separately. 

\subsubsection{$A_1=0$}
\label{exG6A10}

For the branch $A_1=0$, we have $A_0^2=2fX_c$ from 
Eq.~(\ref{A1con}). In this case, the model with 
the sixth-order coupling 
\be
G_6(X)=\sum_{n=2}b_n \left( X-X_c
\right)^n\,,
\label{funcexG6}
\ee
where $X_c=M_{\rm pl}^2$, 
gives rise to the following exact solution with 
the extremal RN metric \cite{GPBH,GPBH2}:
\be
f=h=\left(1-\frac{M}{r}\right)^2\,,\qquad 
A_0=\epsilon \sqrt{2} \Mpl \left(1-\frac{M}{r}\right)\,,\qquad 
A_1=0\,.
\label{sol1exG6}
\ee
Substituting Eqs.~(\ref{funcexG6})-(\ref{sol1exG6}) 
into Eqs.~(\ref{q1def})-(\ref{q2def}), 
(\ref{cr1})-(\ref{F2def}), and (\ref{casq}), 
it follows that the quantities 
$q_1,q_2,{\cal F}_1, {\cal F}_2, c_{r1}^2,c_{r2}^2,
c_{\Omega \pm}^2$ are of the same forms as those in GR given by Eq.~(\ref{GR}) with $f=h=(1-M/r)^2$.
Thus, the exact solution \eqref{sol1exG6} 
exhibits neither ghost nor Laplacian instabilities. 

\subsubsection{$A_0'=0$}
\label{exG6A0con}

The other branch corresponds to $A_0=P={\rm constant}$ 
with $A_1=\epsilon \sqrt{P^2-2fX_c}/f$.
Since all the $G_6$-dependent terms in 
Eqs.~(\ref{be1})-(\ref{be5}) are multiplied by $A_0'$, 
the background solution simply reduces to 
the stealth Schwarzschild solution given by
\be
f=h=1-\frac{2M}{r}\,,\qquad
A_0=P\,,\qquad 
A_1=\epsilon \frac{\sqrt{r(
P^2r-2rX_c+4MX_c)}}
{r-2M}\,,
\label{sol2exG6}
\ee
which arises for arbitrary couplings $G_6(X)$.
For the existence of $A_1$, 
we require that $P^2r-2rX_c+4MX_c>0$. 
Substituting the solution (\ref{sol2exG6}) into Eqs.~(\ref{q1def})-(\ref{q2def}), 
(\ref{cr1})-(\ref{F2def}), and (\ref{casq}), 
we find that $q_1$, ${\cal F}_1$, 
$c_{r1}^2$, and $c_{\Omega+}^2$ are of the same forms as Eq.~(\ref{GR}) with $f=h=1-2M/r$.

On the other hand, the quantities $q_2$, ${\cal F}_2$, $c_{r2}^2$, and $c_{\Omega-}^2$ are affected by 
the sixth-order coupling $G_6(X)$. 
In the vicinity of the event horizon,
we have
\be
q_2=-\frac{P^2G_{6,X}(X_c)}{8M^2(r-2M)^2}
+{\cal O}((r-2M)^{-1})\,,
\qquad 
c_{\Omega-}^2= -\frac{G_6(X_c)+2M^2}{MP^2G_{6,X}(X_c)}(r-2M)+{\cal O}((r-2M)^{2})\,, 
\ee
respectively. 
The conditions $q_2>0$ and $c_{\Omega-}^2 \geq 0$ translate, respectively, to
\ba
& &
G_{6,X}(X_c)<0\,,\label{G6Xre}\\
& &
G_{6}(X_c) \geq -2M^2\,.
\ea
Around the event horizon, 
the condition ${\cal F}_2 \geq 0$ 
corresponds to
\be
[G_6(X_c)-4M^2] [G_6(X_c)-4M^2+2X_cG_{6,X}(X_c)]\geq0\,,
\label{G6cond3} 
\ee
under which $c_{r2}^2=1+{\cal O}(r-2M)$.
Then, for $X_c>0$, the Laplacian instabilities 
are absent under the conditions 
\be
-2M^2 \leq G_6(X_c) \leq 4M^2\qquad {\rm or} \qquad
G_6(X_c)\geq4M^2-2X_cG_{6,X} (X_c)\,.
\label{G6Xre2}
\ee
At spatial infinity ($r\gg 2M$), the quantities 
$q_2$, $c_{r2}^2$, 
and $c_{\Omega-}^2$ reduce to 
those in GR with $f \to 1$ and $h \to 1$.
In summary, the solution (\ref{sol2exG6}) is 
subject to neither ghost nor Laplacian instabilities 
under the conditions (\ref{G6Xre}) and (\ref{G6Xre2}).

\subsection{Other intrinsic vector-mode couplings}
\label{invecsec}

\subsubsection{Coupling $g_4(X)$}

Let us consider the model given by the coupling
\be
G_{2}(X,F)=-2g_4(X) F\,,
\label{g4}
\ee
which was originally introduced in Ref.~\cite{Heisenberg} 
as an intrinsic vector mode in the quartic Lagrangian 
${\cal L}_4$. In this case, Eq.~(\ref{be5}) reduces to 
\be
g_{4,X} A_0'^2 A_1=0\,.
\label{be5exg4}
\ee
For the model 
\be
g_4(X)=g_4(X_c)+\sum_{n=2}b_n 
\left( X-X_c \right)^n\,,
\label{funcexg4}
\ee
which satisfies the condition $g_{4,X}(X_c)=0$, there exists the RN-type exact solution 
\ba
f=h=1-\frac{2M}{r}+\frac{Q^2}{2M_{\rm pl}^2 r^2} 
\left[1-2g_4(X_c) \right]\,,\qquad 
A_0=P+\frac{Q}{r}\,,\qquad 
A_1=\epsilon \frac{\sqrt{A_0^2-2fX_c}}{f}\,.
\label{sol1exg4}
\ea
Plugging the solution (\ref{sol1exg4}) with (\ref{funcexg4}) into Eqs.~(\ref{q2def}) and (\ref{F2def}), it follows that 
\be
q_2=\frac{1-2g_4(X_c)}{2r^2f}\,,\qquad 
{\cal F}_2=\frac{M_{\rm pl}^4 [1-2g_4(X_c)]^2}{16r^8} 
\geq 0\,.
\ee
The other quantities $q_1,{\cal F}_1,c_{r1}^2,
c_{\Omega+}^2,c_{\Omega-}^2$ are of the same 
forms as Eq.~(\ref{GR}) with $f$ and $h$ given by Eq.~(\ref{sol1exg4}). Then, the solution (\ref{sol1exg4}) 
is plagued by neither ghost nor Laplacian instabilities for 
\be
g_4(X_c)< \frac12\,.
\label{g4Xc}
\ee

{}From Eq.~(\ref{be5exg4}), we also have the 
branch characterized by $A_0'=0$.
In this case, the $g_4$-dependent terms in the background equations completely vanish, so the resulting solution 
is the same as Eq.~(\ref{sol2exG6}). 
On using this solution, we find that neither ghost nor 
Laplacian instabilities arise under the 
condition same as Eq.~(\ref{g4Xc}). 

We note that Eq.~(\ref{be5exg4}) admits the 
other branch $A_1=0$. 
For the model (\ref{funcexg4}) with $g_4(X_c)=1/2$, 
there exists the other stealth Schwarzschild solution 
$f=h=1-2M/r$, $A_0=\epsilon \sqrt{2(1-2M/r)X_c}$, 
$A_1=0$. In this case the quantity $q_2$ exactly vanishes, 
so the strong-coupling problem is present.

\subsubsection{Coupling $g_5(X)$}

For the quintic intrinsic vector-mode coupling $g_5(X)$, 
Eq.~(\ref{be5}) gives
\be
\left[ fg_5-\left( A_0^2-2 f X_c \right)g_{5,X} 
\right]A_0'^2=0\,.
\label{g5eq}
\ee

For the branch characterized by 
$fg_5=\left( A_0^2-2 f X_c \right)g_{5,X}$, 
the model satisfying the conditions $g_5(X_c)=0$ 
and $g_{5,X}(X_c)=0$, e.g.,  
\be
g_5(X)=\sum_{n=2}b_n \left( X-X_c 
\right)^n\,,
\label{g5form}
\ee
gives rise to the RN solution (\ref{RNso}) with 
the longitudinal vector component (\ref{A1con}).
In this case, we find that all the quantities in 
Eqs.~(\ref{q1def})-(\ref{q2def}), (\ref{cr1})-(\ref{F2def}), and (\ref{casq}) are the same as 
those in GR given by Eq.~(\ref{GR}).

For the other branch $A_0'=0$ of Eq.~(\ref{g5eq}), 
we obtain the stealth Schwarzschild solution same as Eq.~(\ref{sol2exG6}) for arbitrary couplings $g_5(X)$.
Substituting this solution into Eqs.~(\ref{q1def})-(\ref{q2def}), 
(\ref{cr1})-(\ref{F2def}), and (\ref{casq}), 
the quantities $q_1$, ${\cal F}_1$, 
$c_{r1}^2$, and $c_{\Omega+}^2$ are of the same forms as Eq.~(\ref{GR}) with $f=h=1-2M/r$.
On the other hand, around the horizon, the leading-order 
terms in $q_2$ and $c_{\Omega-}^2$ are given, respectively, by 
\be
q_2=\frac{|P|g_5(X_c)}{4M(r-2M)^{2}}
+{\cal O}((r-2M)^{-1})\,,\qquad 
c_{\Omega-}^2=\frac{M|P|+X_c\,
g_5(X_c)}{MP^2g_5(X_c)}(r-2M)
+{\cal O}((r-2M)^{2})\,.
\ee
Then, the conditions $q_2>0$ and 
$c_{\Omega-}^2 \geq 0$ translate to
\ba
& &
g_5(X_c)>0\,,
\label{g5Cond0}\\
& &
M|P|+X_c\,g_5(X_c) \geq 0\,, 
\label{g5Cond1}
\ea
respectively. The condition ${\cal F}_2 \geq 0$ 
corresponds to 
\be
\left[M-|P|g_5(X_c) \right]
\left[
M
-\frac{P^2-X_c}{|P|}
g_5(X_c)
\right] \geq 0\,,
\label{g5Cond2}
\ee
under which $c_{r2}^2=1+{\cal O}(r-2M)$. 
There exists the parameter space in which all
the conditions (\ref{g5Cond0})-(\ref{g5Cond2}) 
are satisfied.
At spatial infinity ($r\gg 2M$), 
the quantities $q_2$, $c_{r2}^2$, 
and $c_{\Omega-}^2$ are the same as those in GR. 

\section{Quartic power-law couplings}
\label{sec5}

In Sec.~\ref{exG4}, we studied whether or not 
the conditions for the absence of ghost and Laplacian instabilities are satisfied for some exact BH solutions 
by imposing the two conditions (\ref{excon}). 
Now, without imposing the conditions (\ref{excon}),
we extend the analysis to more general quartic 
power-law couplings given by 
\be
G_4(X)=\frac{\Mpl^2}{2}+\beta_4\Mpl^2 
\left(\frac{X}{\Mpl^2}\right)^n\,,
\label{modelG4}
\ee
where $\beta_4$ and $n~(\geq 1)$ are constants.
{}From Eq.~(\ref{be5exG4}), there are several branches 
characterized by $A_1 \neq 0$ or $A_1=0$. 
For $n \geq 3$, there is a branch satisfying 
$A_1^2=A_0^2/(fh)$ besides other 
branches discussed below. 
This corresponds to the RN solution obeying the 
specific relation $X=0$ from Eq.~(\ref{Xdef}). 
In this case we have $G_4=\Mpl^2/2$ and 
$G_{4,X}=G_{4,XX}=0$, so the quantities 
$q_1,q_2,{\cal F}_1,{\cal F}_2,c_{r1}^2,c_{r2}^2,
c_{\Omega \pm}^2$ trivially reduce to those in GR 
given by Eq.~(\ref{GR}). 
In the following, we will focus on other branches 
of Eq.~(\ref{be5exG4}).

\subsection{Branch with $A_1\neq0$}

Let us consider the quartic power-law model with $n \geq 2$.
Solving Eq.~(\ref{be5exG4}) for $A_1$, there is the 
branch satisfying 
\be
A_1=\pm \sqrt{\frac{A_0^2 [f(h-1)+(2n-1)rf'h]
-4rA_0A_0'(n-1)fh}{fh[(2n-1)rf'h-(1+h-2nh)f]}}\,.
\label{G4n2A1}
\ee
The effect of coupling $\beta_4$ on $f$ and $h$
appears as corrections to the RN metric 
expressed in the form 
\be
f_{\rm RN}=h_{\rm RN}=
\left( 1-\frac{r_h}{r} \right)
\left( 1- \mu \frac{r_h}{r} \right)\,,
\label{fRN}
\ee
where $\mu$ is a constant 
in the range $0<\mu<1$.
In the vicinity of the event horizon characterized by 
the distance $r_h$, we expand $f,h,A_0$ as follows: 
\be
f=\sum_{i=1}^{\infty} f_i(r-r_h)^i\,,\qquad
h=\sum_{i=1}^{\infty} h_i(r-r_h)^i\,,\qquad
A_0=a_0+\sum_{i=1}^{\infty} a_i(r-r_h)^i\,,
\label{fhA0hor}
\ee
where $f_i,h_i,a_0,a_i$ are constants. 
We assume that $a_0>0$ without loss of generality.
Substituting Eq.~(\ref{fhA0hor}) into 
Eqs. (\ref{be1})-(\ref{be5}) and solving them iteratively, 
the coefficients $f_1, h_1, a_1$ are known to be 
\be
f_1=h_1=\frac{1-\mu}{r_h}>0\,,\qquad 
a_1=\frac{\sqrt{2\mu}M_{\rm pl}}{r_h}
+\tilde{a}_1\,,
\ee
where $\tilde{a}_1$ is a $\beta_4$-dependent constant that vanishes in the limit $\beta_4 \to 0$ \cite{GPBH,GPBH2}.
The coupling $\beta_4$ arises in metric components 
at the orders of $i \geq 2$.
To study the stability of hairy BH solutions against odd-mode perturbations for the model
(\ref{modelG4}), we do not need 
the explicit expressions of coefficients in Eq.~(\ref{fhA0hor}). 
Plugging Eq.~(\ref{fhA0hor}) into Eq.~(\ref{G4n2A1}), 
the longitudinal component around the event horizon behaves as 
\be
A_1=\frac{a_0}{f_1(r-r_h)}
+\frac{a_0[f_2+h_2-2f_1^2(n-1)-f_1(f_2+h_2)(2n-1)r_h]
+2a_1f_1(f_1r_h-1)}{2f_1^2[(2n-1)f_1r_h-1]}
+{\cal O}(r-r_h)\,,
\label{A1exrh}
\ee
which diverges as $r \to r_h$. 
This divergent property also persists for the exact solution discussed in Sec.~\ref{exG4}, 
but the regularity of the coordinate-independent scalar quantity $\int A_{\mu}dx^{\mu}$ is 
ensured by introducing advanced and retarded null 
coordinates \cite{GPBH,GPBH2}.
We recall that, due to the property $D_2=0$ for general quartic couplings $G_4(X)$, 
the two solutions of Eq.~(\ref{casq}) 
reduce to $c_{\Omega+}^2={C_1r^2}/(q_1f)$ and $c_{\Omega-}^2=-{D_1r^2}/(q_2f)$. 
Substituting Eq.~(\ref{fhA0hor}) into Eqs.~(\ref{q1def}), (\ref{q2def}), and $c_{\Omega+}^2$, and then picking up 
the leading-order contributions around the event horizon, 
the product $q_1q_2c_{\Omega+}^2$ yields
\be
q_1q_2c_{\Omega+}^2= 
-\left[\frac{(n-1)a_0(a_0+2a_1r_h)}{\Mpl^2\{(2n-1)f_1r_h-1\}}\right]^{2(n-1)}
\frac{n^2a_0^2\beta_4^2}{4f_1^3r_h^2(r-r_h)^{3}}
+{\cal O}\left((r-r_h)^{-2}\right)\,, 
\label{q1q2casqG4}
\ee
which is always negative for $\beta_4 \neq 0$. 
We note that the charged stealth Schwarzschild solution 
discussed in Sec.~\ref{exG4A1neq0}, i.e., $n=1$ and $\beta_4=1/4$ in Eq.~(\ref{modelG4}), also has the same property. Setting $n=1$, $\beta_4=1/4$, $r_h=2M$, 
$f_1=1/(2M)$, and $a_0=P+Q/(2M)$ in Eq.~(\ref{q1q2casqG4}), 
we reproduce the result 
given by Eq.~(\ref{qcpro}).

{}From Eq.~(\ref{q1q2casqG4}), the conditions for 
the absence of ghosts ($q_1>0$, $q_2>0$) and those 
for no Laplacian instability along the 
angular direction ($c_{\Omega+}^2 \geq 0$) cannot be 
simultaneously satisfied. 
Even if $\beta_4$ is very close to 0, the first term on 
the right hand side of Eq.~(\ref{q1q2casqG4}) dominates 
over the other terms in the limit that $r \to r_h$. 
This means that, no matter how small the coupling 
constant $\beta_4$ is, the hairy BH solutions present 
for the quartic model (\ref{modelG4}) are unstable 
in the vicinity of the event horizon.
This instability is mostly attributed to the existence 
of nonvanishing longitudinal component $A_1$ 
which exhibits the divergence at $r=r_h$.

\subsection{Branch with $A_1=0$}

Let us proceed to the other branch characterized by $A_1=0$ 
for the power-law model (\ref{modelG4}) with $n \geq 1$.
For this branch, the coefficients $C_2$ and $C_5$ in 
Eq.~(\ref{q2def}) simply reduce to $C_2=0$ and $C_5=1/(2fr^2)$, respectively, so that 
\be
q_2=\frac{1}{2fr^2}\,.
\ee
Then, the second no-ghost condition (\ref{NG2}) is always 
satisfied throughout the horizon exterior.

The coupling $\beta_4$ works as corrections to 
the RN metric given by Eq.~(\ref{fRN}). 
Plugging Eq.~(\ref{fhA0hor}) into 
Eqs.~(\ref{be1})-(\ref{be4}), the resulting iterative 
solution around the event horizon has the 
leading-order terms:
\be
f_1=h_1=\frac{1-\mu}{r_h}>0\,,\qquad 
a_0=0\,,\qquad a_1=\frac{\sqrt{2\mu}M_{\rm pl}}
{r_h}\,,
\label{f1h1}
\ee
so that the temporal component $A_0$ vanishes 
on the event horizon.
The coupling $\beta_4$ appears 
in the coefficients of expansions 
of $f, h, A_0$ at the order of 
$(r-r_h)^{n+1}$ \cite{GPBH,GPBH2}.
Substituting the iterative solution (\ref{fhA0hor}) with 
(\ref{f1h1}) into Eq.~(\ref{q1def}),
we obtain
\be
q_1=\frac{\Mpl^2}{4f_1 (r-r_h)}+{\cal O}\left((r-r_h)^{0}\right)\,,
\ee
which means that the first no-ghost condition 
(\ref{NG1}) is satisfied. 
Around the event horizon, the quantities (\ref{F1def}) and (\ref{F2def}) 
yield 
\be
{\cal F}_1=\frac{M_{\rm pl}^4 f_1^2}{4r_h^8} 
(r-r_h)^2+{\cal O}\left((r-r_h)^{3}\right)\,,\qquad 
{\cal F}_2=\frac{M_{\rm pl}^4}{16r_h^8}
+{\cal O}\left(r-r_h\right)\,,
\label{F12form}
\ee
which are both positive, as required 
for the absence of Laplacian instabilities 
in the radial direction.

For general quartic couplings $G_4(X)$ 
with the branch $A_1=0$, 
the propagation speeds along the radial and angular 
directions have the following relations:
\ba
c_{r1}^2
&=&c_{\Omega+}^2=
\frac{fG_4}{fG_4-A_0^2G_{4,X}}\,,
\label{crG41} \\
c_{r2}^2
&=&c_{\Omega-}^2=
\frac{fG_4+A_0^2G_{4,X}(G_{4,X}-1)}
{fG_4-A_0^2G_{4,X}}\,.
\label{crG42}
\ea
On using the expanded solution (\ref{fhA0hor}) around 
the event horizon with Eq.~(\ref{f1h1}), 
Eqs.~(\ref{crG41}) and (\ref{crG42}) reduce to 
\ba
c_{r1}^2
&=&c_{\Omega+}^2=
1+\beta_4 \frac{n}{2^{n-2}}
\left( \frac{2\mu}{1-\mu} \right)^n
\left( \frac{r}{r_h}-1 \right)^n
+{\cal O}\left((r-r_h)^{n+1}\right)\,,\\
c_{r2}^2
&=&c_{\Omega-}^2=
1+\beta_4^2 \frac{n^2}{2^{2n-3}}
\left( \frac{2\mu}{1-\mu} \right)^{2n-1}
\left( \frac{r}{r_h}-1 \right)^{2n-1}
+{\cal O}\left((r-r_h)^{2n}\right)\,,
\ea
which approach 1 in the limit that $r \to r_h$. 
Hence the Laplacian instabilities are absent around 
$r=r_h$. 
The small deviations of $c_{r1}^2$ and $c_{r2}^2$ from 1 
arise from the coupling $\beta_4$ in the vicinity of the event horizon.

At spatial infinity ($r \gg r_h$), the solutions of $f,h,A_0$ 
expanded as a series of $1/r$ are given by 
\ba
f&=&1-\frac{2M}{r} \left[1 
-\frac{2\beta_4 n (2MM_{\rm pl}\tilde{P}+\sqrt{2}Q)
\tilde{P}^{2n}}
{MM_{\rm pl}\tilde{P}\{2\beta_4 (2n-1) \tilde{P}^{2n}
-1\}} \right]+\frac{1}{r^2} 
\left(\frac{Q^2}{2M_{\rm pl}^2}+\mu_1 \right)
+{\cal O}(r^{-3})\,,
\label{fla}\\
h&=&1-\frac{2M}{r}+\frac{1}{r^2} 
\left( \frac{Q^2}{2M_{\rm pl}^2}+\mu_2 \right)+{\cal O}(r^{-3})\,,\\
A_0&=&P+\frac{Q}{r}+\frac{\mu_3}{r^2}+{\cal O}(r^{-3})\,,
\label{A0la}
\ea
where $\tilde{P}=P/(\sqrt{2}M_{\rm pl})$, and 
$\mu_1,\mu_2,\mu_3$ are $r$-independent constants 
containing the dependence of $\beta_4$ 
(which vanish in the limit that $\beta_4 \to 0$). 
The terms $\mu_1,\mu_2,\mu_3$ do not appear for 
the dominant contributions to 
$q_1,q_2,c_{r1}^2,c_{r2}^2,c_{\Omega+}^2,
c_{\Omega-}^2$.
The leading-order term of $q_1$ is given by 
\be
q_1=\frac{\Mpl^2}{4}\frac{\left[1-2(2n-1)\beta_4\tilde{P}^{2n}\right]^2}
{1+2\beta_4\tilde{P}^{2n}}\,,
\label{qG4inf}
\ee
so the ghost instabilities are absent for 
\be
\beta_4 \tilde{P}^{2n}>-\frac12\,.
\label{noghostG4l}
\ee
The leading-order contributions to the quantities 
(\ref{F1def}) and (\ref{F2def}) are 
\ba
{\cal F}_1 
&=& \frac{M_{\rm pl}^4}{4r^8} 
\left(1+2\beta_4 \tilde{P}^{2n}
\right) \left[ 1-2(2n-1)\beta_4\tilde{P}^{2n}\right]\,,\\
{\cal F}_2
&=& \frac{M_{\rm pl}^4}{16r^8}
\left[ 1-2(2n-1)\beta_4\tilde{P}^{2n}\right]
\left[ 1-2(2n-1)\beta_4\tilde{P}^{2n}
+4n^2 \beta_4^2 \tilde{P}^{4n-2}\right]\,.
\ea
Under the condition (\ref{noghostG4l}), 
the two conditions ${\cal F}_1 \geq 0$ and 
${\cal F}_2 \geq 0$ are satisfied for 
\be
1-2(2n-1)\beta_4\tilde{P}^{2n} \geq 0 
\qquad
{\rm and}\qquad
1-2(2n-1)\beta_4\tilde{P}^{2n}
+4n^2 \beta_4^2 \tilde{P}^{4n-2} \geq 0\,.
\label{nolapG4l}
\ee
The leading-order sound speeds are given 
by\footnote{For the model (\ref{modelG4}) with $A_1\ne0$, 
the quantities $q_1,q_2$ and $c_{r1}^2,c_{r2}^2,c_{\Omega+}^2,c_{\Omega-}^2$ at spatial infinity also reduce to Eqs.~(\ref{qG4inf}) and (\ref{crG4inf}) at leading order, respectively, by reflecting the fact that 
$A_1$ vanishes in the limit that $r \to \infty$ \cite{GPBH,GPBH2}.}
\be
c_{r1}^2=c_{\Omega+}^2=1+\frac{4n\beta_4\tilde{P}^{2n}}{1-2(2n-1)\beta_4\tilde{P}^{2n}}\,,\qquad 
c_{r2}^2=c_{\Omega-}^2=1+\frac{4n^2\beta_4^2\tilde{P}^{4n-2}}{1-2(2n-1)\beta_4\tilde{P}^{2n}}\,,
\label{crG4inf}
\ee
which are both positive under the inequalities
(\ref{noghostG4l}) and (\ref{nolapG4l}).
Hence there are neither ghost nor Laplacian instabilities 
under the conditions (\ref{noghostG4l}) and (\ref{nolapG4l}), which are well satisfied for 
$|\beta_4|$ and $|\tilde{P}|$ smaller than the order of unity.

\begin{figure}
\begin{center}
\includegraphics[height=3.4in,width=3.4in]{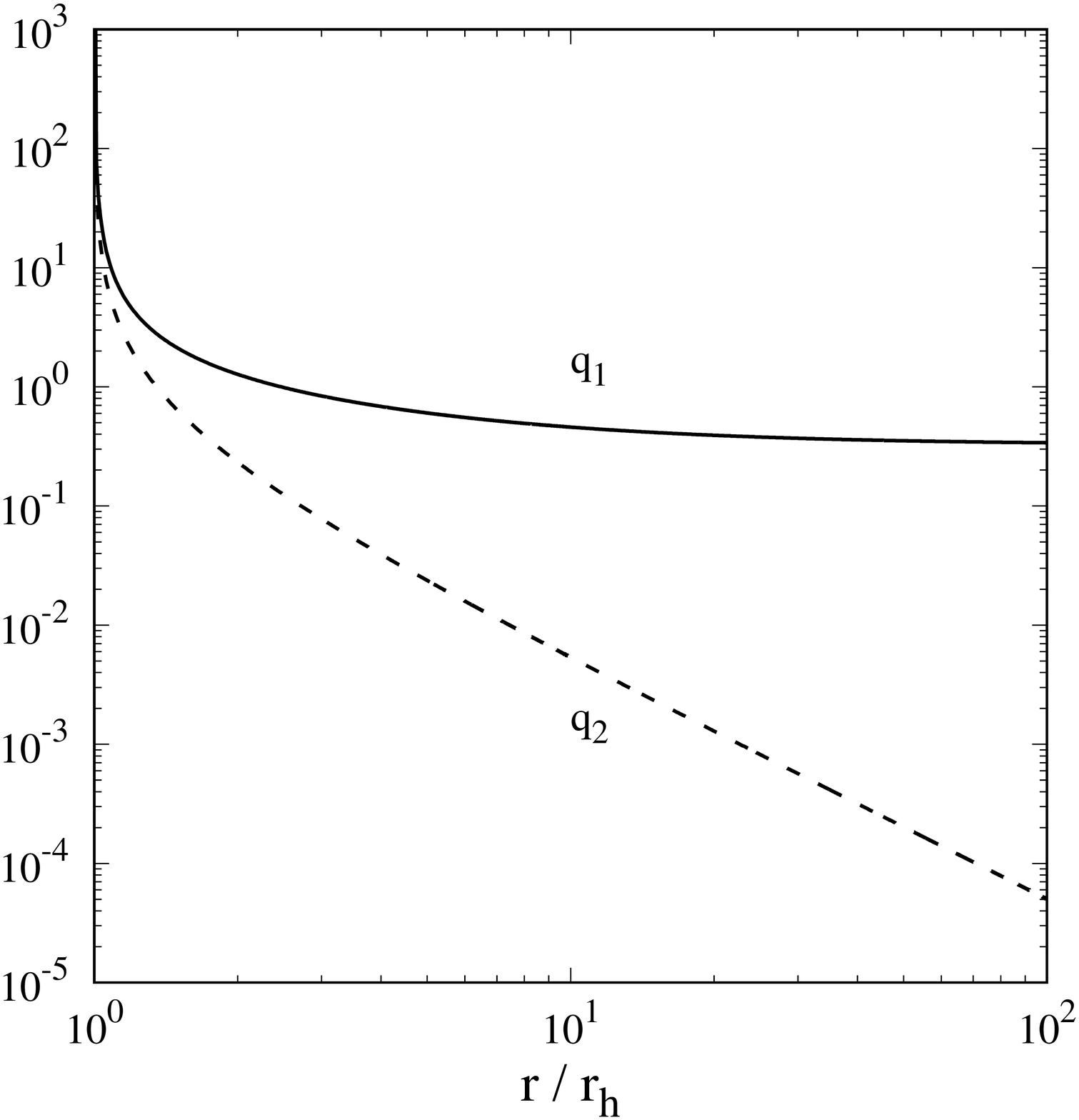}
\includegraphics[height=3.4in,width=3.4in]{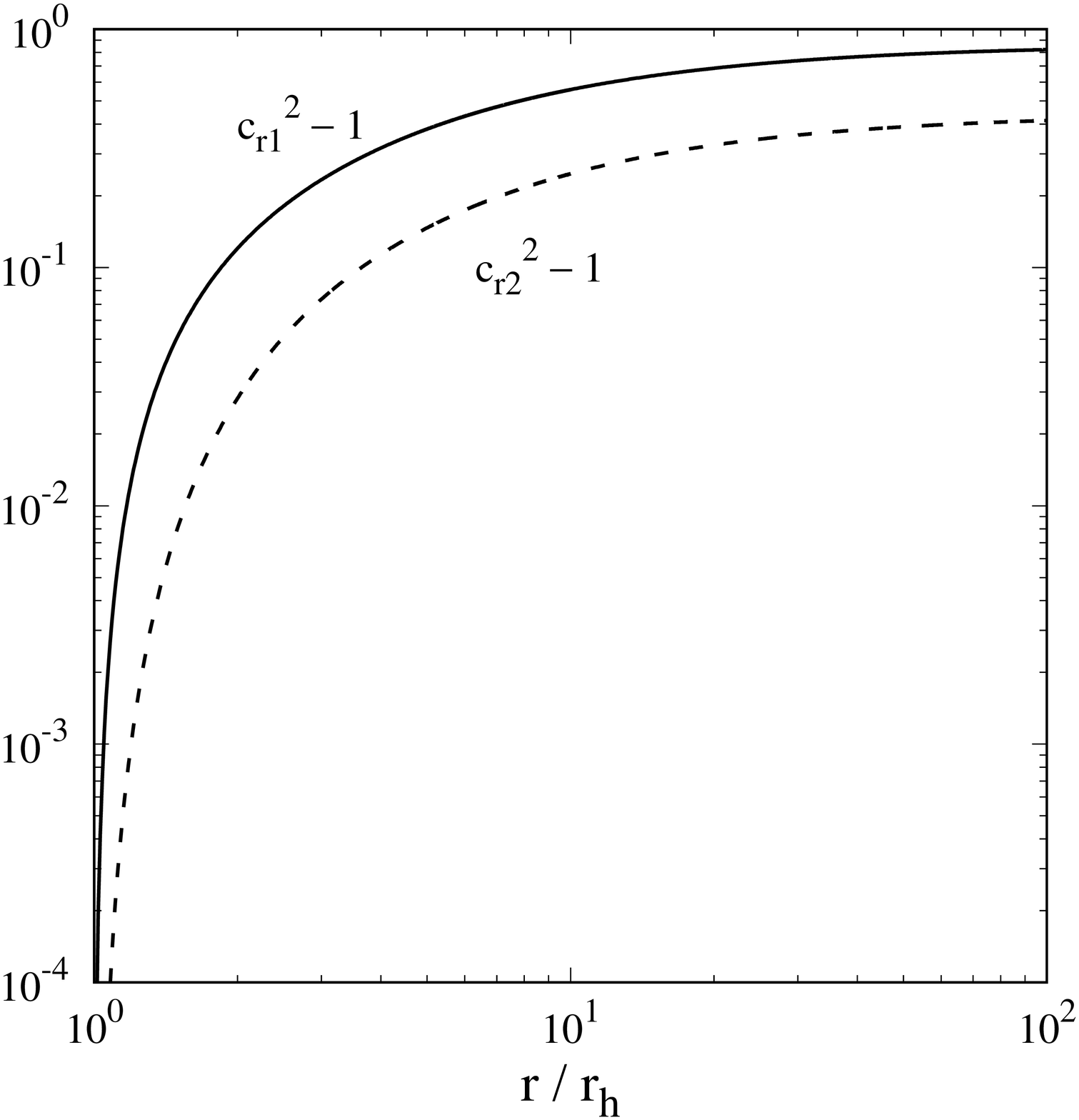}
\end{center}
\caption{\label{fig1}
Numerical plots of $q_1$ and $q_2$ normalized by 
$M_{\rm pl}^2/4$ and $r_h^{-2}$ respectively (left), 
and the deviations of
$c_{r1}^2, c_{r2}^2$ from 1 (right) for the quartic 
power-law model (\ref{modelG4}) with $n=2$ 
and $A_1=0$. We choose the parameters as 
$\beta_4=1$, $\tilde{P}=0.5$, and $\mu=0.2$. 
}
\end{figure}

To confirm the above analytic estimations, we numerically compute
the radial dependence of $q_1, q_2$ as well as $c_{r1}^2, c_{r2}^2$ 
for hairy BH solutions present in the model (\ref{modelG4}). 
For the branch $A_1=0$, we recall that $c_{\Omega+}^2$ and 
$c_{\Omega-}^2$ are equivalent to $c_{r1}^2$ and $c_{r2}^2$, 
respectively.
We numerically integrate Eqs.~(\ref{be1})-(\ref{be4}) 
for $n=2$ by employing the boundary conditions of $f,h,A_0$ 
around the event horizon (see equations (6.17)-(6.19) in Ref.~\cite{GPBH2}) 
and substitute the background solutions into 
(\ref{q1def})-(\ref{q2def}) and (\ref{cr1})-(\ref{cr2}).
In Fig.~\ref{fig1}, we plot $q_1,q_2$ and $c_{r1}^2-1, c_{r2}^2-1$ 
versus $r/r_h$ for $\beta_4=1$ and $\tilde{P}=0.5$. 
In this case, the conditions  (\ref{noghostG4l}) and (\ref{nolapG4l}) are consistently satisfied at spatial infinity.
In Fig.~\ref{fig1}, we can confirm that there are neither ghost 
nor Laplacian instabilities throughout the horizon exterior. 
The asymptotic values of $c_{r1}^2$ and $c_{r2}^2$ on the event horizon are equivalent to 1, while at spatial infinity, $c_{r1}^2$ and $c_{r2}^2$ approach 
constants different from 1. This difference is induced by 
the non-zero coupling $\beta_4$. 

\section{Sixth-order power-law couplings}
\label{sec6}

In this section, we consider the sixth-order power-law couplings 
given by 
\be
G_6(X)=\frac{\beta_6}{\Mpl^2}\left(\frac{X}{\Mpl^2}\right)^n\,, 
\label{modelG6}
\ee
with $G_4=\Mpl^2/2$, where $\beta_6$ and 
$n~(\geq 0)$ are constants.
In this case, the longitudinal mode obeys
\be
\beta_6 A_0'^2A_1 \left( A_0^2 -fh A_1^2 \right)^{n-2}
\left[ A_1^2 fh \{ (2n+1)h-1 \}-A_0^2 (3h-1) 
\right]=0\,. 
\label{G6A1eq}
\ee
{}From Eq.~(\ref{G6A1eq}), there are four possible 
branches characterized by 
(i) $A_0'=0$, (ii) $A_1=0$, (iii) $A_0^2=fhA_1^2$ 
(present for $n \geq 3$), and 
(iv) $A_1^2=A_0^2(3h-1)/[fh\{(2n+1)h-1\}]$ \cite{GPBH,GPBH2}. 
The branches (i) and (iii) give rise to the stealth 
Schwarzschild solution with $A_1$ undetermined 
and the trivial RN solution, respectively. 
The branch (iv) does not exist throughout the horizon 
exterior ($0<h<1$).
Then, we focus on the branch (ii), i.e.,  
\be
A_1=0\,.
\ee

The theory with $n=0$ corresponds to the $U(1)$-gauge 
invariant vector-field interaction advocated 
by Horndeski \cite{Horndeski76}, in which case 
the hairy BH solution with $A_1=0$ was found
in Ref.~\cite{HorndeskiBH}. 
In this case, we have $G_{6,X}=0$, $C_9=0$ and 
$C_8=-fhC_{10}$, so the quantity (\ref{F1def}) 
reduces to ${\cal F}_1=4f^3hC_{10}^2 \geq 0$.
{}From Eq.~(\ref{cr1}), it follows that 
\be
c_{r1}^2=1\qquad ({\rm for}~n=0)\,.
\ee
For the power-law models with $n \geq 0$, let us study 
whether the theoretical consistent conditions derived in 
Sec.~\ref{sec3} are satisfied
by using the iterative solutions 
(\ref{fhA0hor}) in the vicinity of the event horizon.  
The coupling $\beta_6$ appears as 
corrections to the RN solution (\ref{fRN}).
For $n=0$, the coefficients of Eq.~(\ref{fhA0hor}) consistent with the background Eqs.~(\ref{be1})-(\ref{be4}) are
\ba
\hspace{-0.3cm}
&&
f_1=h_1=\frac{1-\mu}{r_h}\,,\qquad a_1=
\frac{M_{\rm pl}}{r_h}\sqrt{\frac{2\mu}
{1+2\tilde{\beta}_6}}\,,\nonumber \\
\hspace{-0.3cm}
&&
f_2=-\frac{1-2\mu+(2-5\mu+3\mu^2)\tilde{\beta}_6}
{(1+2\tilde{\beta}_6)r_h^2}\,,\qquad
h_2=-\frac{1-2\mu+(2-\mu-\mu^2)\tilde{\beta}_6}
{(1+2\tilde{\beta}_6)r_h^2}\,,\qquad 
a_2=-\frac{M_{\rm pl}}{r_h^2} \sqrt{\frac{2\mu}
{1+2\tilde{\beta}_6}}
\frac{1-\tilde{\beta}_6}{1+2\tilde{\beta}_6}\,,
\label{a2def}
\ea
where $\tilde{\beta_6} \equiv \beta_6/(r_h^2\Mpl^2)$,
$\mu$ is a constant appearing in the RN metric 
(\ref{fRN}), and $a_0 \neq 0$.
Substituting these iterative solutions into Eqs.~(\ref{q1def}) and (\ref{q2def}), we obtain
\ba
q_1=\frac{\Mpl^2(1+2\tilde{\beta}_6)r_h}{4(1-\mu)
[1+2(1-\mu)\tilde{\beta}_6](r-r_h)}+{\cal O}\left((r-r_h)^{0}\right)\,,\qquad 
q_2= \frac{1-(1-\mu)\tilde{\beta}_6}
{2(1-\mu)r_h(r-r_h)}+{\cal O}\left((r-r_h)^{0}\right)\,.
\ea
The leading-order terms of 
$c_{r2}^2, c_{\Omega+}^2,c_{\Omega-}^2$ 
in the vicinity of the event horizon yield
\be
c_{r2}^2=1\,,\qquad
c_{\Omega+}^2=
\frac{1+2(1-\mu)\tilde{\beta}_6}{1+2\tilde{\beta}_6}\,,
\qquad 
c_{\Omega-}^2=
\frac{[1+2(1-\mu)\tilde{\beta}_6]^2}{(1+2\tilde{\beta}_6)[1-(1-\mu)\tilde{\beta}_6]}\,.
\ee
If the coupling $\tilde{\beta}_6$ is in the range
\be
-\frac12<\tilde{\beta}_6<\frac{1}{1-\mu} 
\qquad ({\rm for}~n=0)\,,
\label{be6con}
\ee
the conditions $q_1>0,q_2>0,c_{\Omega+}^2 \geq 0, 
c_{\Omega-}^2 \geq 0$ are satisfied. 
Since ${\cal F}_2=\Mpl^4[1-(1-\mu)\tilde{\beta}_6]^2/(16r_h^8)$, the condition (\ref{NL2}) is automatically 
satisfied.

For the models with $n \geq 1$, the coefficients in the expansions (\ref{fhA0hor})
up to the order of $r-r_h$ are given by 
\be
f_1=h_1=\frac{1-\mu}{r_h}\,,\qquad
a_0=0\,,\qquad a_1=\frac{\sqrt{2\mu}M_{\rm pl}}{r_h}\,.
\label{f1h1co}
\ee
The coupling $\beta_6$ appears for the terms 
whose orders are higher than $r-r_h$. 
Then, the quantities $q_1$ and 
$q_2$ reduce, respectively, to 
\be
q_1\simeq \frac{\Mpl^2}{4f_1(r-r_h)}
+{\cal O}\left((r-r_h)^{0}\right)\,,\qquad 
q_2\simeq \frac{1}{2f_1r_h^2(r-r_h)}
+{\cal O}\left((r-r_h)^{0}\right)\,,
\ee
so that the two conditions (\ref{NG1}) 
and (\ref{NG2}) are satisfied. 
The quantities ${\cal F}_1$ and ${\cal F}_2$ are the 
same as those given in Eq.~(\ref{F12form}), so the 
conditions (\ref{NL1}) and (\ref{NL2}) are also satisfied. 
For $n=1$, the propagation speeds are given by 
\be
c_{r1}^2=1+{\cal O}(r-r_h)\,,\qquad 
c_{r2}^2=1+{\cal O}\left((r-r_h)^2\right)\,,\qquad 
c_{\Omega+}^2=1+{\cal O}(r-r_h)\,,\qquad 
c_{\Omega-}^2=1-\mu \tilde{\beta}_6+{\cal O}(r-r_h)\,.
\ee
Then, there are neither ghost nor Laplacian instabilities 
under the condition 
\be
\tilde{\beta}_6 \leq \frac{1}{\mu}\qquad 
({\rm for}~n=1)\,.
\label{be6co}
\ee
If $n \geq 2$, then the propagation speeds yield
\be
c_{r1}^2=1+{\cal O}((r-r_h)^n)\,,\qquad 
c_{r2}^2=1+{\cal O}\left((r-r_h)^{2n}\right)\,,\qquad 
c_{\Omega+}^2=1+{\cal O}((r-r_h)^n)\,,\qquad 
c_{\Omega-}^2=1+{\cal O}((r-r_h)^{n-1})\,,
\label{nge2pro}
\ee
where we used the property $f_i=h_i$ for $i \leq n$ in the 
expansions of Eq.~(\ref{fhA0hor}) \cite{GPBH2}.
The coupling $\beta_6$ gives rise to the deviations of 
$c_{r1}^2,c_{r2}^2,c_{\Omega+}^2,c_{\Omega-}^2$ from 1.
{}From Eq.~(\ref{nge2pro}), the Laplacian instabilities are absent 
in the vicinity of the event horizon.

\begin{figure}
\begin{center}
\includegraphics[height=3.4in,width=3.4in]{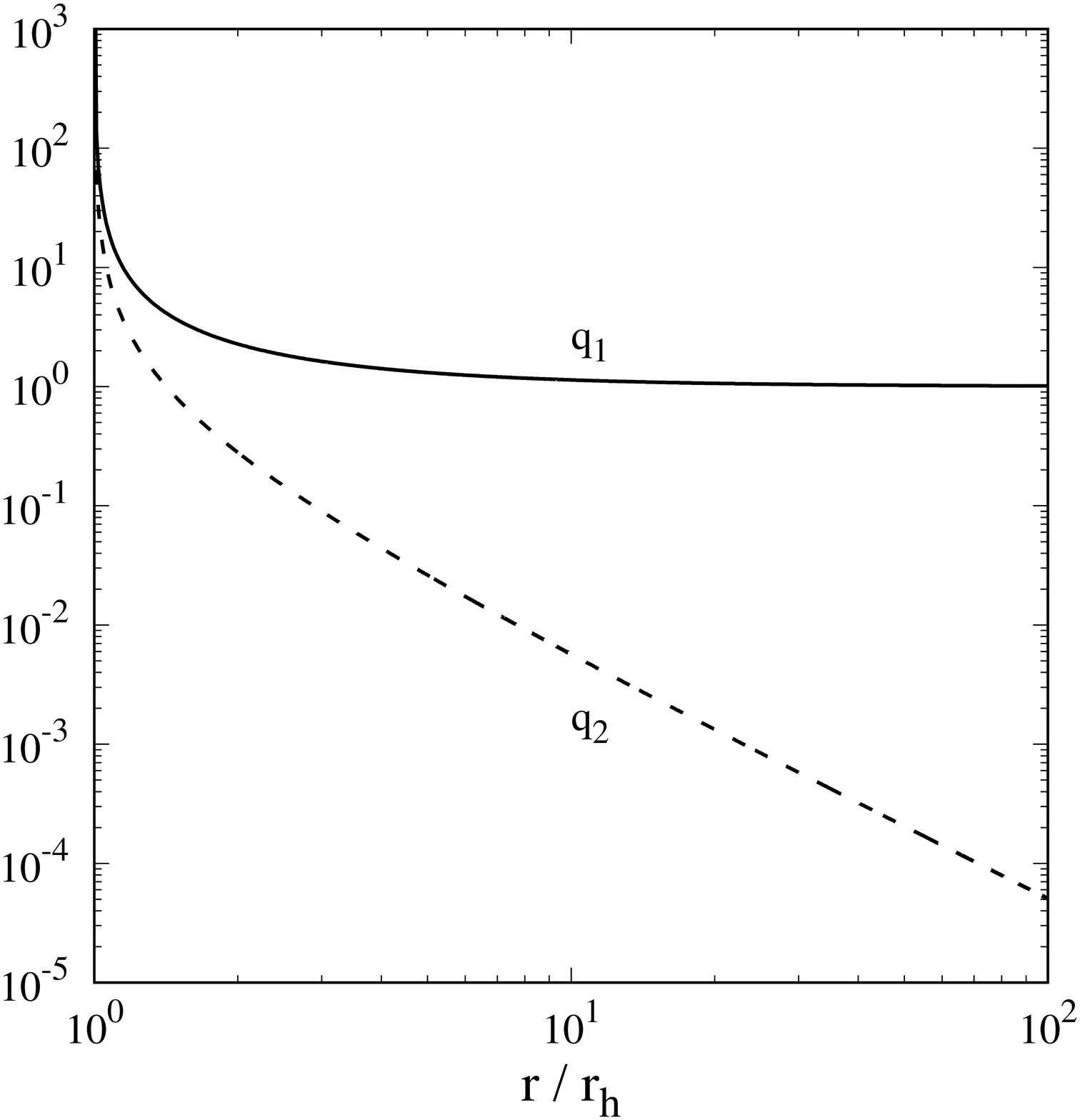}
\includegraphics[height=3.4in,width=3.4in]{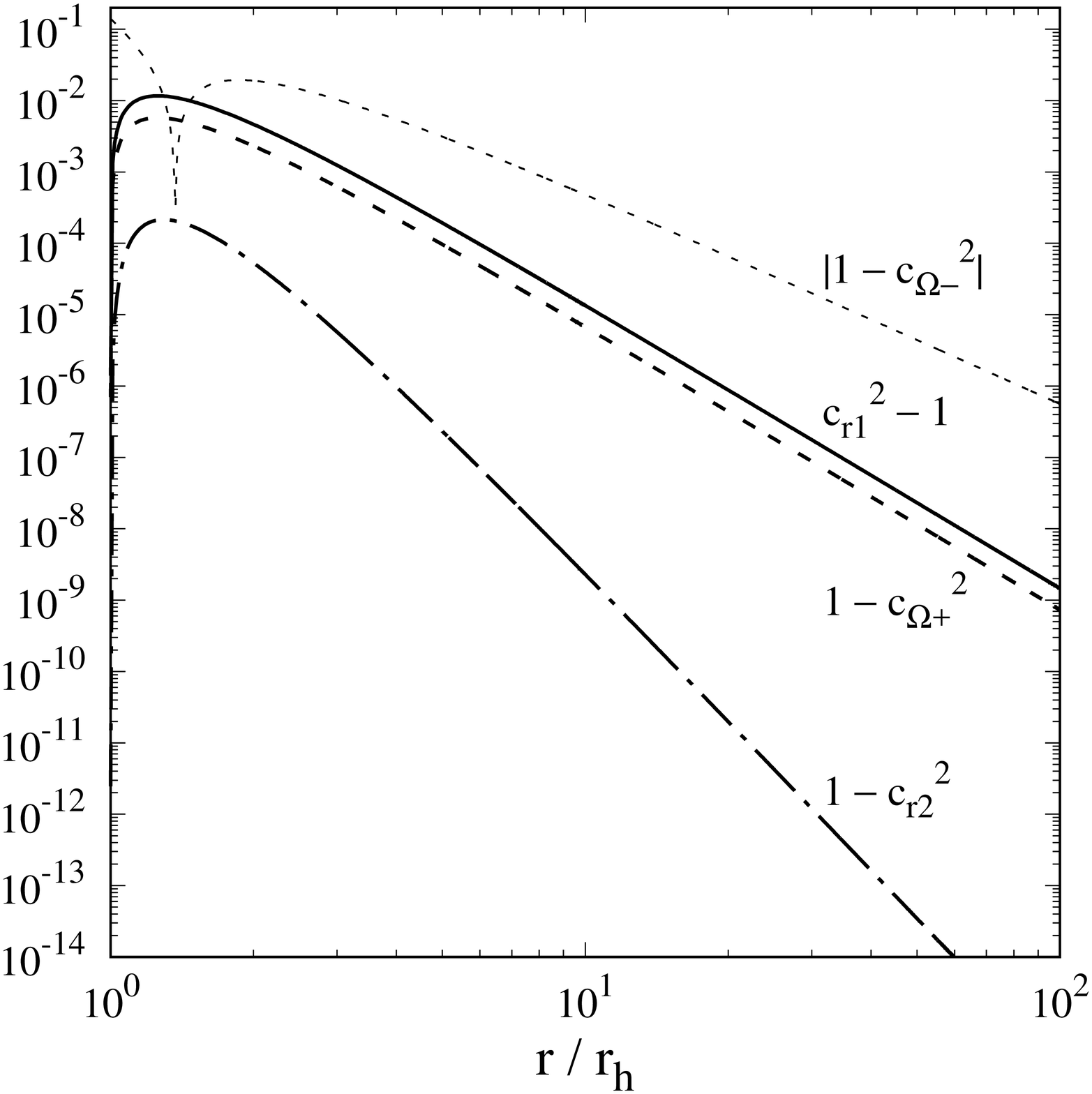}
\end{center}
\caption{\label{fig2}
Numerical plots of $q_1$ and $q_2$ normalized by 
$M_{\rm pl}^2/4$ and $r_h^{-2}$ respectively (left), 
and the deviations of 
$c_{r1}^2$, $c_{r2}^2$, $c_{\Omega+}^2$, $c_{\Omega-}^2$ from 1 (right) 
for the sixth-order power-law model 
(\ref{modelG6}) with $n=1$. 
The model parameters are chosen to be 
$\tilde{\beta}_6=0.7$ and $\mu=0.2$. 
}
\end{figure}

At spatial infinity ($r\gg r_h$), the solutions of 
$f,h,A_0$ expanded as a series of $1/r$ are 
given \cite{GPBH,GPBH2} by
\ba
f&=& 1-\frac{2M}{r}+\frac{Q^2}{2M_{\rm pl}^2 r^2}
-\frac{\beta_6 P^{2n}Q^2}
{2^{1+n}M_{\rm pl}^{4+2n}r^4}
-\frac{2^{-n}\beta_6 P^{2n-1}Q^2 
\left[M P (6n-5)+8Qn \right]}
{10M_{\rm pl}^{4+2n}r^5} +{\cal O}\left( 
r^{-6} \right)\,,\label{f0G6}\\
h&=& 1-\frac{2M}{r}+\frac{Q^2}{2M_{\rm pl}^2 r^2}
+\frac{\beta_6 M P^{2n} Q^2(2n-1)}{2^{1+n}M_{\rm pl}^{4+2n}r^5}
+{\cal O}\left( 
r^{-6} \right)\,,\\
A_0&=&
P+\frac{Q}{r}-\frac{2^{-n}\beta_6 M P^{2n}Q}
{M_{\rm pl}^{2+2n}r^4}
-\frac{2^{-n}\beta_6 P^{2n-1}Q
(32M^2M_{\rm pl}^2P n+28MM_{\rm pl}^2 Q\,n
-3P Q^2)}{20M_{\rm pl}^{4+2n}r^5}+{\cal O}\left( 
r^{-6} \right)\,,
\label{A0P}
\ea
which means that the coupling $\beta_6$ works as 
corrections to the RN solution (\ref{RNso}). 
On using Eqs.~(\ref{f0G6})-(\ref{A0P}), 
it follows that $q_1,q_2,{\cal F}_1,{\cal F}_2$ 
approach the GR values (\ref{GR}) with $f=h=1$. 
The propagation speed squares 
$c_{r1}^2,c_{r2}^2,c_{\Omega+}^2,c_{\Omega-}^2$ 
also approach the asymptotic value 1.
For instance, the deviation of $c_{r1}^2$ from 1 
far outside the event horizon behaves as 
\be
c_{r1}^2-1\simeq 
2n\tilde{\beta}_6\tilde{P}^{2n}\tilde{Q}^2\left(\frac{r_h}{r}\right)^4\,, 
\label{G6cr1}
\ee
where $\tilde{P}=P/(\sqrt{2}M_{\rm pl})$ and 
$\tilde{Q}=Q/(r_h\Mpl)$. 
Then, $c_{r1}^2-1$ decreases rapidly for increasing $r$.

By integrating Eqs.~(\ref{be1})-(\ref{be4}) with respect to $r$ 
with the boundary conditions of $f,h,A_0$ around the event horizon, we can numerically obtain hairy BH solutions 
induced by the coupling $\beta_6$. 
For such background solutions,  
we compute the quantities associated with the absence of 
ghost and Laplacian instabilities. 
In Fig.~\ref{fig2}, we plot $q_1,q_2$ as well as 
the deviations of 
$c_{r1}^2,c_{r2}^2,c_{\Omega+}^2,c_{\Omega-}^2$ 
from 1 as functions of $r/r_h$ for the model parameters 
$n=1$, $\tilde{\beta}_6=0.7$, and $\mu=0.2$.
In this case, the condition (\ref{be6co}) 
is satisfied around the event horizon. 
As we see in the left panel of Fig.~\ref{fig2}, both 
$q_1$ and $q_2$ are positive throughout the horizon 
exterior. In the right panel, we observe that  
$c_{r1}^2,c_{r2}^2,c_{\Omega+}^2$ approach 1 
as $r \to r_h$, while $c_{\Omega-}^2 \to 
1-\mu \tilde{\beta}_6$ in the same limit. 
The propagation speed squares show some deviations from 1 
slightly outside the event horizon, but they rapidly approach the 
asymptotic value 1 for the distance $r \gg r_h$.
Thus, there are neither ghost nor Laplacian instabilities 
throughout the horizon exterior. 
Under the condition (\ref{be6con}), this property also holds 
for the BH solution arising from the $U(1)$ gauge-invariant interaction ($n=0$).

\section{Other intrinsic vector-mode couplings}
\label{sec7}

Finally, we study whether or not hairy BH solutions arising from intrinsic vector modes $g_4(X)$ and $g_5(X)$ satisfy 
the conditions for the absence of ghosts and Laplacian 
instabilities. We will mostly discuss the power-law models $g_4(X) \propto X^n$ and $g_5(X) \propto X^n$, but as we see below,  
it is possible to derive several conditions in a general way without restricting their functional forms.

\subsection{Coupling $g_4(X)$}

Let us consider the coupling $G_2(X,F)=-2g_4(X)F$ 
with $G_4=M_{\rm pl}^2/2$.
In this case, the longitudinal 
mode $A_1$ satisfies the relation (\ref{be5exg4}).
Independent of the branches arising from Eq.~(\ref{be5exg4}), 
the quantities $q_1,q_2, c_{r1}^2,c_{r2}^2,c_{\Omega+}^2,c_{\Omega-}^2$ are given, respectively, by 
\be
q_1=\frac{\Mpl^2 h}{4f^2}\,,\qquad 
q_2=\frac{1-2g_4}{2r^2f}\,,\qquad 
c_{r1}^2=c_{r2}^2=c_{\Omega+}^2=c_{\Omega-}^2=1\,,
\label{q1q2g4}
\ee
with ${\cal F}_1=M_{\rm pl}^4fh/(4r^8) \geq 0$ and 
${\cal F}_2=M_{\rm pl}^4 (1-2g_4)^2h^3/(16r^8f^3) \geq 0$.
Hence there are neither ghost nor Laplacian instabilities 
under the condition  
\be
g_4<\frac{1}{2}\,.
\label{g4con}
\ee

Let us consider the power-law coupling model given by  
\be
g_4(X)=\gamma_4\left(\frac{X}{\Mpl^2}\right)^n\,, 
\label{modelg4}
\ee
where $\gamma_4$ and $n~(\geq 1)$ are constants. 
{}From Eq.~(\ref{be5exg4}), there are three branches characterized by (i) $A_0'=0$, (ii) $A_1=0$, and 
(iii) $A_1^2=A_0^2/(fh)$ (present for $n\geq2$). 
The branches (i) and (iii) correspond to 
the stealth Schwarzschild solution and the RN solution, respectively. 
For the branch (ii), there exists a hairy BH solution 
where the coupling $\gamma_4$ appears in 
$f,h,A_0$ as corrections to the RN 
solution \cite{GPBH,GPBH2}. 

For the branch $A_1=0$, the coupling (\ref{modelg4}) 
reduces to 
\be
g_4=\gamma_4\left(\frac{A_0^2}{2f\Mpl^2}\right)^n\,.
\label{q2g4}
\ee
Around the event horizon, the leading-order solutions to $f,h,A_0$ 
are given by $f\simeq h \simeq (1-\mu)(r-r_h)/r_h$ 
and $A_0 \simeq \sqrt{2\mu}\Mpl (r-r_h)/r_h$, respectively, with the corrections of coupling $\gamma_4$ 
appearing at the order 
of $(r-r_h)^{n+1}$  \cite{GPBH,GPBH2}. 
Then, the term (\ref{q2g4}) vanishes in the limit $r\to r_h$, 
so the ghost instability is absent around the event horizon.
At spatial infinity ($r\gg r_h$), the solutions behave as 
$f,h\to1$ and $A_0\to P={\rm constant}$, 
so the coupling (\ref{q2g4}) reduces to 
$g_4 \simeq \gamma_4 \tilde{P}^{2n}$, where 
$\tilde{P}=P/(\sqrt{2}M_{\rm pl})$. 
In this regime, the condition (\ref{g4con}) translates to 
$\gamma_4 \tilde{P}^{2n}<1/2$.
For $n \geq 1$, the temporal vector component 
squared $A_0^2$ monotonically increases from 
0 ($r\simeq r_h$) to the 
asymptotic value $P^2$ ($r\to\infty$). 
Hence, under the condition 
$\gamma_4 \tilde{P}^{2n}<1/2$, there is no ghost instability 
throughout the horizon exterior.

\subsection{Coupling $g_5(X)$}

We proceed to study the quintic intrinsic vector-mode coupling $g_5(X)$ with $G_4=\Mpl^2/2$. 
Independent of the functional form of $g_5$, we have 
the following relations:
\be
q_1=\frac{h\Mpl^2}{4f^2}\,,\qquad 
c_{r1}^2=1\,.
\label{g5q1cr1}
\ee
On the other hand, the quantities 
$q_2,c_{r2}^2,c_{\Omega+}^2, 
c_{\Omega-}^2$ depend on the choice of $g_5(X)$.

In what follows, we focus on the power-law coupling 
model given by 
\be
g_5(X)=\frac{\gamma_5}{\Mpl^2}\left(\frac{X}{\Mpl^2}\right)^n\,, 
\label{modelg5}
\ee
where $\gamma_5$ and $n~(\geq 1)$ are constants. 
{}From Eq.~(\ref{g5eq}), there are three branches characterized by $A_0'^2=0$, 
$A_1=\epsilon \sqrt{A_0^2/(fh)}$ (present for 
$n \geq 2$), and 
\be
A_1=\epsilon \sqrt{\frac{A_0^2}{(1+2n)fh}}\,.
\label{A1g5}
\ee
The first two branches correspond to the stealth Schwarzschild solution and the RN solution, 
respectively \cite{GPBH2}. 
For the branch (\ref{A1g5}), there exists a hairy BH solution 
where the coupling $\gamma_5$ works as corrections 
to the RN metric, so we investigate its stability 
in the following.

For $n \geq 1$, the solutions expanded around $r=r_h$ 
are given by Eq.~(\ref{fhA0hor}) with the 
coefficients:
\be
f_1=h_1=\frac{1-\mu}{r_h}\,,\qquad
f_2=h_2=\frac{2\mu-1}{r_h^2}\,,\qquad
a_0=0\,,\qquad a_1=\frac{\sqrt{2\mu}M_{\rm pl}}{r_h}\,,
\qquad a_2=-\frac{\sqrt{2\mu}M_{\rm pl}}{r_h^2}\,,
\ee
up to the order of $(r-r_h)^2$.
The coupling $\gamma_5$ appears at the order of 
$(r-r_h)^{n+2}$ in the expansions of $f,h,A_0$ \cite{GPBH2}. 
On using this iterative solution, 
it follows that 
\be
q_2=\frac{1}{2f_1r_h^2(r-r_h)}+{\cal O}((r-r_h)^{0})\,,
\qquad
{\cal F}_2=\frac{M_{\rm pl}^4}{16r_h^8}
+{\cal O}(r-r_h)\,,
\ee
so the conditions (\ref{NG2}) and (\ref{NL2}) 
are satisfied around $r=r_h$. 
For $\gamma_5>0$, the propagation speed squares yield
\be
c_{r2}^2=1+{\cal O}((r-r_h)^{n+1})\,,\qquad
c_{\Omega+}^2=1+{\cal O}((r-r_h)^{n+1})\,,\qquad
c_{\Omega-}^2=1+{\cal O}((r-r_h)^{n})\,,
\ee
whereas, for $\gamma_5<0$, the power-law dependence between 
$c_{\Omega+}^2-1$ and $c_{\Omega-}^2-1$ is exchanged. 
The coupling $\gamma_5$ induces the deviations of 
$c_{r2}^2, c_{\Omega+}^2, c_{\Omega-}^2$ from 1.
Thus, the hairy BH solution is free from the Laplacian 
instability in the vicinity of the event horizon.

For the distance $r \gg r_h$, the coupling $\gamma_5$ 
gives rise to corrections to the RN solution (\ref{RNso}) at the 
order of $1/r^3$ in $f,h$ and at the order of $1/r^2$ 
in $A_0$ \cite{GPBH2}. 
{}From Eq.~(\ref{A1g5}), the longitudinal mode approaches 
a constant $A_1 \to P/\sqrt{1+2n}$ as $r \to \infty$.
On using the iterative solution at spatial infinity, we find 
that $q_2 \simeq 1/(2r^2)>0$ and 
${\cal F}_2=M_{\rm pl}^4/(16r^8)>0$. 
Moreover, the propagation speed squares 
$c_{r2}^2, c_{\Omega+}^2, c_{\Omega-}^2$ approach 1 
as $r \to \infty$, so the Laplacian instability is absent at 
spatial infinity.
In the intermediate regime between $r=r_h$ and $r \gg r_h$, 
we numerically confirmed that the hairy BH solutions arising 
from the power-law coupling with $n \geq 1$ are plagued 
by neither ghost nor Laplacian instabilities.
The coupling $\gamma_5$ gives rise to the values of $c_{r2}^2, c_{\Omega+}^2, c_{\Omega-}^2$ different from 1 in the 
intermediate region, but they rapidly approach 1 for increasing
$r$. For example, the quantity $c_{r2}^2-1$ has 
the dependence proportional to $1/r^2$ at spatial infinity.
The qualitative behavior of  $c_{r2}^2, c_{\Omega+}^2, c_{\Omega-}^2$
 is similar to that shown in the right 
panel of Fig.~\ref{fig2}.

\section{Conclusions}
\label{consec}

In this paper, we formulated the odd-parity 
perturbations about the static and spherically symmetric
BH solutions in generalized Proca theories
by expanding the action up to the second order 
in perturbations. We derived the conditions 
under which hairy BH solutions in these theories
suffer neither ghost nor Laplacian instabilities.
The existence of a temporal vector component $A_0$ besides 
a longitudinal mode gives rise to a bunch of exact and numerical BH solutions 
with vector hairs.

For odd-parity perturbations, there are two propagating degrees of freedom 
arising from the gravity sector and the vector field. 
For $l \geq 2$, where $l$ is an integer associated with the expansion 
in terms of spherical harmonics $Y_{lm}$, 
the conditions for avoiding ghost instabilities 
correspond to $q_1>0$ and $q_2>0$, 
where $q_1,q_2$ are given by Eqs.~(\ref{q1def})-(\ref{q2def}).  
We derived the propagation speeds in the radial direction 
in the forms (\ref{cr1})-(\ref{cr2}), 
where the quantities ${\cal F}_1$ and ${\cal F}_2$ 
are required to be positive for the existence of real 
solutions of $c_{r1}$ and $c_{r2}$. In the angular direction, there are also two propagation speed 
squares $c_{\Omega \pm}^2$ derived in Eq.~(\ref{casq}), which must be 
positive to avoid Laplacian instabilities for large $l$.
We also showed that the analysis of dipole perturbations ($l=1$) does not give rise to additional conditions to those obtained for $l \geq 2$. 
For the cubic couplings $G_3(X)$, there are no ghost and Laplacian instabilities against odd-parity perturbations 
as in GR.

The charged stealth Schwarzschild BH solution (\ref{sol1exG4}) with nonvanishing $A_1$ 
present for the quartic coupling 
(\ref{func1exG4}) satisfies neither ${\cal F}_1 \ge 0$ nor
$c_{\Omega+}^2 \geq 0$ in the vicinity of the event horizon 
under the no-ghost conditions (\ref{noghostTa}). 
Then, this solution, 
which was firstly found in Ref.~\cite{Tasinato1} 
for the model $G_4(X)=M_{\rm pl}^2/2+X/4$, 
is prone to the Laplacian instability. 
For models with the quartic coupling (\ref{func2exG4}) and 
with the quintic coupling (\ref{funcexG5}), 
there exist the extremal RN solution (\ref{sol2exG4}) with $A_1=0$ and the RN solution with $A_1 \neq 0$, 
respectively.
These two exact BH solutions are plagued by neither 
ghost nor Laplacian instabilities. 
Models with intrinsic vector modes, e.g., 
(\ref{funcexG6}), (\ref{funcexg4}), and (\ref{g5form}), also give rise to exact BH solutions satisfying the relations (\ref{excon}). In such cases, we showed the existence of parameter spaces consistent 
with conditions for the absence of ghost and Laplacian instabilities.

For the quartic power-law models (\ref{modelG4}), there are hairy BH solutions where the coupling $\beta_4$ appears as corrections to the RN solution (\ref{RNso}).
For the branch $A_1 \neq 0$, we can iteratively derive the solutions to $f,h,A_0$ in the forms (\ref{fhA0hor}) 
around the event horizon.
On using this expansion, we showed that the product $q_1q_2c_{\Omega+}^2$ is given by Eq.~(\ref{q1q2casqG4}), which is negative 
for $\beta_4 \neq 0$. 
Even if $\beta_4$ is very small, the first term on the right hand side of Eq.~(\ref{q1q2casqG4}) 
dominates over the other terms as $r$ approaches $r_h$. 
Then, the hairy BH solutions present for the power-law 
model (\ref{modelG4}) with $A_1 \neq 0$ 
are prone to the instability problem in the vicinity 
of the event horizon for $\beta_4\neq 0$ and $n \geq 1$.
For the model (\ref{modelG4}) with the branch $A_1=0$, 
the BH solution suffers neither ghost nor Laplacian instabilities throughout the horizon exterior 
under the conditions (\ref{noghostG4l}) and (\ref{nolapG4l}). 
This suggests that the nonvanishing longitudinal mode $A_1$ 
with peculiar behavior (\ref{A1exrh}) around $r=r_h$ is the main 
reason for the instability mentioned above. 

We also investigated the case of hairy BH solutions with the branch $A_1=0$ arising from the sixth-order power-law coupling (\ref{modelG6}).
The model with $n=0$, which corresponds to the $U(1)$ gauge-invariant 
derivative interaction, has the iterative BH solution (\ref{fhA0hor}) 
with the coefficients given by (\ref{a2def}) and $a_0 \neq 0$. 
In this case, we showed that ghost and Laplacian instabilities 
are absent under the condition (\ref{be6con}). 
For $n=1$, the coupling $\beta_6$ leads to a nontrivial deviation of 
$c_{\Omega-}^2$ from 1 around $r=r_h$, 
so that the Laplacian instability can be avoided under the condition 
(\ref{be6co}). For $n \geq 2$, there are no particular bounds on 
the coupling $\beta_6$. In this case, 
$c_{r1}^2,c_{r2}^2,c_{\Omega+}^2,c_{\Omega-}^2$ approach 1 
in both limits $r \to r_h$ and $r \to \infty$, with 
small deviations from 1 in the intermediate regime
outside the event horizon. 

For general intrinsic vector-mode couplings $g_4(X)$, there are neither ghost nor Laplacian instabilities for $g_4<1/2$.
In case of the power-law couplings (\ref{modelg4}), 
the hairy BH solution with the branch $A_1=0$ has the property $g_4=0$ on the event horizon, 
so it is sufficient to satisfy the condition $g_4<1/2$ at spatial infinity. 
For the quintic intrinsic vector-mode power-law 
couplings (\ref{modelg5}) with $n \geq 1$, 
the conditions for the absence of 
ghost and Laplacian instabilities are satisfied throughout the 
horizon exterior without a particular bound on 
the coupling $\gamma_5$.

In summary, we showed that, apart from the quartic 
power-law models (\ref{modelG4}) with the branch 
$A_1 \neq 0$, there are hairy BH solutions 
which are free from ghost and Laplacian instabilities
against the odd-parity perturbations. 
In Tables \ref{tab1} and \ref{tab2} in Appendix \ref{appenD}, we summarize the stability of BH solutions found in 
the analysis of Secs.~\ref{sec4}-\ref{sec6}.
We should emphasize, however, that the analysis of odd-parity perturbations alone is not sufficient to guarantee the complete stability of BHs in general.
We need to ensure whether the BH solutions 
satisfy the same type of stability conditions 
against even-parity perturbations.
Moreover, even if the absence of ghost and Laplacian instabilities for both odd- and even-parity sectors is 
proven for some BH solutions, 
we still need to ensure the absence of
tachyonic eigenmodes for the radial perturbation 
equations in both sectors 
(See Ref.~\cite{GGGP} for related arguments in 
Horndeski theories).
Finally, it also has to be checked
that all the solutions to the linear and nonlinear perturbation equations
obtained from regular initial data
remain bounded throughout the horizon exterior \cite{Kay,Dafermos}. 

The other important question is whether or not 
the BH solutions free from ghost and Laplacian 
instabilities can be further constrained 
from the observational points of view. 
The almost simultaneous detection of GWs from 
a NS merger and its short gamma-ray burst counterpart
has significantly constrained the deviation of propagation speed of GWs from the speed of light 
to be less than order $10^{-15}$ \cite{LIGO2}.
For example,
if we naively apply this bound to 
the large-distance modification of the propagation speeds of perturbations \eqref{crG4inf} 
derived from the $A_1=0$ branch of the quartic power-law
coupling models \eqref{modelG4},
we would obtain a bound $|\beta_4 {\tilde P}^{2n}|\lesssim 10^{-15}$,
assuming that $|\beta_4 {\tilde P}^{2n}|\ll 1$ and $n=O(1)$.
However, we have to be more careful 
when we relate the propagation speeds 
derived in this paper to the observed speed 
of GWs from local sources (BHs and NSs). 

First, the observed polarized GWs $h_+$ and $h_{\times}$ 
are the combination of even- and odd-parity perturbations in general \cite{Martel}, so we need to 
consider the propagation of even modes
for deriving the speed of GWs appropriately. 
Second, the GWs recently detected by LIGO and Virgo from 
a NS merger travelled over the cosmological distance \cite{LIGO2}, so strictly speaking,
it is required to consider the propagation of GWs 
on the time-dependent background. It will be of interest to derive the GW propagation speed 
from local sources by taking into account even-parity
perturbations as well as the effect of time-dependent cosmological background.
These issues will be left for future work.

\appendix

\section{
Coefficients in the background equations}
\label{appa}

In Eqs.~(\ref{be1})-(\ref{be3}) the coefficients 
$c_{1,2,\cdots,19}$ are given by 
\bea
c_{1} &=& -A_{1} X G_{3,X},
\\
c_{2} &=& -2 G_{4} + 4 (X_{0} + 2 X_{1}) G_{4,X} + 8 X_{1} X G_{4,XX},
\\
c_{3} &=& -A_{1} (3 h X_{0} + 5 h X_{1} - X) G_{5,X} - 2 h A_{1} X_{1} X G_{5,XX},
\\
c_{4} &=& G_{2} - 2 X_{0} G_{2,X} - \frac{h}{f} (A_{0} A_{1} A_{0}' + 2 f X A_{1}') G_{3,X} 
-\frac{h A'^{2}_{0}(1+2 G_{2,F})}{2 f}\,,
\\
c_{5} &=& -4 h A_{1} X_{0} G_{3,X} - 4 h^{2} A_{1} A'_{1} G_{4,X} 
+ \frac{8 h}{f} \left( A_{0} X_{1} A'_{0} - f h A_{1} X A'_{1} \right) G_{4,XX}
+\frac{2 h^{2}}{f} A_{1} A'^{2}_{0} (g_{5} + 2 X_{0} g_{5,X}),
\\
c_{6} &=& 2 (1 - h) G_{4} + 4 (h X - X_{0}) G_{4,X} +8 h X_{0} X_{1} G_{4,XX} 
- \frac{h}{f} \left[ (h - 1) A_{0} A_{1} A'_{0} + 2 f (3 h X_{1} + h X_{0} - X) 
A'_{1} \right] G_{5,X}
\\&&-\frac{2 h^{2} X_{1}}{f} (A_{0} A_{1} A'_{0} + 2 f X A'_{1}) G_{5,XX} 
+ \frac{h A'^{2}_{0}}{f} \left[ (h - 1) G_{6} + 2 (h X - X_{0}) G_{6,X} + 4 h X_{0} X_{1} G_{6,XX} \right],
\\
c_{7} &=& -G_{2} + 2 X_{1} G_{2,X} - \frac{h}{f} A_{0} A_{1} A'_{0} G_{3,X}
+\frac{h A'^{2}_{0}(1+2 G_{2,F})}{2 f}\,,
\\
c_{8} &=& 4 h A_{1} X_{1} G_{3,X} + \frac{4 h}{f} A_{0} A'_{0} (G_{4,X}+2 X_{1} G_{4,XX})
-\frac{2 h^{2}}{f} A_{1} A'^{2}_{0} (3 g_{5} + 2 X_{1} g_{5,X})\,, 
\\
c_{9} &=& 2 (h - 1) G_{4} - 4 (2 h - 1) X_{1} G_{4,X} - 8 h X_{1}^{2} G_{4,XX} 
- \frac{h}{f} A_{0} A_{1} A'_{0}\left[ (3 h -1) G_{5,X} + 2 h X_{1} G_{5,XX} \right] 
\\&&-\frac{h}{f} A'^{2}_{0} \left[ (3 h - 1) G_{6} + 2 (6 h - 1) X_{1} G_{6,X} + 4 h X_{1}^{2} G_{6,XX} \right],
\\
c_{10} &=& -\frac{2h}{f} (G_{4} - 2 X G_{4,X})\,,
\\
c_{11} &=& -\frac{2h^{2}}{f} A_{1} X G_{5,X}\,,
\\
c_{12} &=& \frac{h}{f^2} [G_{4} - 2 (2 X_{0} + X_{1}) G_{4,X} - 4 X_{0} X G_{4,XX}]\,,
\\
c_{13} &=& \frac{h^{2}}{f^2} A_{1} [(3 X_{0} + X_{1}) G_{5,X} + 2 X_{0} X G_{5,XX}]\,,
\\
c_{14} &=& - \frac{h}{f} A_{1} [(3 X_{0} + 5 X_{1}) G_{5,X} + 2 X_{1} X G_{5,XX}]\,,
\\
c_{15} &=& \frac{h}{f^{2}}
[ 2 f A_{1} X_{0} G_{3,X} + 2 (2 A_{0} A'_{0} - f h A_{1} A'_{1}) G_{4,X}
+ 4 \left\{ A_{0} (2 X_{0} + X_{1}) A'_{0} - f h A_{1} X A'_{1} \right\} G_{4,XX} \nonumber \\
& &-h A_{1} A'^{2}_{0} (g_{5} + 2 X_{0} g_{5,X})],
\\
c_{16} &=& -\frac{h}{f^{2}}\left[2 f (G_{4} - 2 X G_{4,X} + 4 X_{0} X_{1} G_{4,XX}) 
+ h \left\{ 3 A_{0} A_{1} A'_{0} + 2 f (X_{0} + 3 X_{1}) A'_{1} \right\} G_{5,X}
\right.
\\ && \left.
+2 h \left\{ A_{0} A_{1} (X_{1} + 2 X_{0}) A'_{0} + 2 f X_{1} X A'_{1} \right\} G_{5,XX}
+ h A'^{2}_{0} (G_{6} + 2 X G_{6,X} + 4 X_{0} X_{1} G_{6,XX}) \right],
\\
c_{17} &=& - 2 G_{4} + 8 X_{1} (G_{4,X} + X_{1} G_{4,XX}) 
\\ && + \frac{h A'_{0}}{f} \left[ A_{0} A_{1} (3 G_{5,X} + 2 X_{1} G_{5,XX}) 
+ A'_{0} \left\{ 3 G_{6} + 4 X_{1} (3 G_{6,X} + X_{1} G_{6,XX}) \right\} \right], 
\\
c_{18} &=& 2 G_{2} - \frac{2 h}{f} \left[ (A_{0} A_{1} A'_{0} + 2 f X_{1} A'_{1}) G_{3,X}
+ 2 (A_{0} A''_{0} + A'^{2}_{0}) G_{4,X}
+ 2 A'_{0} (2 X_{0} A'_{0} - h A_{0} A_{1} A'_{1}) G_{4,XX} \right]
\\ &&+ \frac{2 h^{2} A'_{0}}{f^{2}} \left[ f (2 A_{1} A''_{0} + A'_{0} A'_{1}) g_{5}
+ A'_{0} (A_{0} A_{1} A'_{0} + 2 f X_{1} A'_{1}) g_{5,X} \right] 
+\frac{h}{f} A'^{2}_{0}\,,
\\
c_{19} &=& \frac{2 h}{f} \left[ - 2 (A_{0} A'_{0} + f h A_{1} A'_{1}) G_{4,X} 
+ 4 X_{1} (A_{0} A'_{0} - f h A_{1} A'_{1}) G_{4,XX} 
+ h (A_{1} A'^{2}_{0} + A_{0} A'_{0} A'_{1} + A_{0} A_{1} A''_{0}) G_{5,X} 
\right.
\\ && 
+ 2 h A'_{0} (A_{0} X_{1} A'_{1} + A_{1} X_{0} A'_{0}) G_{5,XX}
+ 2 h A'_{0} A''_{0} G_{6}
+ \frac{ h A'_{0}}{f} \left\{ (A_{0} A'^{2}_{0} + 4 f X_{1} A''_{0} - 3 f h A_{1} A'_{0} A'_{1}) G_{6,X}
\right.
\\ && \left.\left.
+ 2 A'_{0} X_{1} (A_{0} A'_{0} - f h A_{1} A'_{1}) G_{6,XX} \right\} \right]\,.
\eea

\section{
Integral of spherical harmonics}
\label{appb}

As we mentioned in the main text, it is sufficient to set 
$m=0$ for the integrations of the second-order action 
of odd-parity perturbations. 
In doing so, we use the following properties:
\ba
& &
\int_0^{2\pi} d\varphi \int_0^{\pi} d\theta\, 
\left| \frac{\partial}{\partial \theta} Y_{l0} 
(\theta, \varphi) \right|^2 \sin \theta=L\,,\\
& &
\int_0^{2\pi} d\varphi \int_0^{\pi} d\theta\, 
\left[ \frac{1}{\sin \theta} \left| \frac{\partial}
{\partial \theta} Y_{l0} 
(\theta, \varphi) \right|^2+\sin \theta \left| 
\frac{\partial^2}{\partial \theta^2} Y_{l0} 
(\theta, \varphi)\right|^2 \right]=L^2\,,
\ea
where $L$ is defined by Eq.~(\ref{Ldef}).

\section{
Coefficients in the second-order action of odd-parity perturbations}
\label{appc}

The coefficients $C_{1-13}$ in Eq.~\eqref{oddLag}
are given by
\bea
&&
C_1=\frac{h(rG_4-2rXG_{4,X}+hA_1XG_{5,X})}{2r^3f}\,,
\label{defC1}\\
&&
C_2=-\frac{h[2rfA_1G_{4,X}-fhA_1^2G_{5,X}+A_0A_0'(2hA_1G_{6,X}-2rg_5)]}{4r^3f^2}\,,
\label{defC2}\\ 
&&
C_3=\frac{h}{2r^3f}\left[rA_0G_{4,X}-\frac12hA_0A_1G_{5,X}
-A_0'\left(hG_6-h^2A_1^2G_{6,X}+rhA_1g_5\right)\right]\,,
\label{defC3}\\ 
&&
C_4=-\frac{h}{2r^4f}\left[A_0'\left\{r^2(1+G_{2,F}-G_{4,X})-h\left(2G_6-\frac12rA_1G_{5,X}
+4rA_1g_5\right)+2h^2A_1^2G_{6,X}\right\}\right.\notag\\
&&\hspace{2.2cm}\left.
+2A_0\left(rG_{4,X}-\frac12hA_1G_{5,X}\right)\right]\,,
\label{defC4}\\ 
&&
C_5=\frac{1}{2r^3f}\left[r(1+G_{2,F})-h'(G_6-hA_1^2G_{6,X})+2h^2A_1A_1'G_{6,X}
-(2hA_1+rh'A_1+2rhA_1')g_5\right]\,,
\label{defC5}\\ 
&&
C_6=\frac{h(fA_0'-f'A_0)(rg_5-hA_1G_{6,X})}{r^3f^2}\,,
\label{defC6}\\ 
&&
C_7=-\frac{h}{2r^3f}\left[rf(1+G_{2,F})-f'h(G_6-hA_1^2G_{6,X})-h(rf'A_1+2fA_1)g_5\right]\,,
\label{defC7}\\ 
&&
C_8=-\frac{h}{4r^4f}\left[2fG_4+2fhA_1^2G_{4,X}+hA_1(f'X-A_0A_0')G_{5,X}
-hA_0'^2(G_6-hA_1^2G_{6,X})\right]\,,
\label{defC8}\\ 
&&
C_9=\frac{h\left[2fA_1G_{4,X}+(f'X-A_0A_0')G_{5,X}+hA_0'^2A_1G_{6,X}\right]}{2r^4f}\,,
\label{defC9}\\ 
&&
C_{10}=\frac{1}{4r^4f^3}\left[2f^2G_4-2fA_0^2G_{4,X}+f\{hA_0A_0'A_1+fX(h'A_1+2hA_1')\}G_{5,X}
-hA_0'^2(fG_6+A_0^2G_{6,X})\right]\,,
\label{defC10}\\ 
&&
C_{11}=-\frac{1}{4r^4f^3}\left[4f^2A_0G_{4,X}-fA_0(f'hA_1+fh'A_1+2fhA_1')G_{5,X}
+2f(f'hA_0'-fh'A_0'-2fhA_0'') G_6\right.\notag\\
&&\hspace{2.4cm}\left.
+2hA_0'(f'A_0^2-fA_0A_0'+f^2h'A_1^2+2f^2hA_1A_1')G_{6,X}\right]\,,
\label{defC11}\\ 
&&
C_{12}=-\frac{1}{4r^4f^3}\left[2f^3(1+G_{2,F})-f(2ff''h-f'^2h+ff'h')G_6
+f'h(f'A_0^2-2fA_0A_0'+f^2h'A_1^2+2f^2hA_1A_1')G_{6,X}\right.\notag\\
&&\hspace{2.4cm}\left.
-2f^2(f'hA_1+fh'A_1+2fhA_1')g_5\right]\,,
\label{defC12}\\ 
&&
C_{13}=-\frac{1}{8r^4f^3}\left[
4r^2f^3G_{2,X}+2rf^2\left(4fhA_1+rfh'A_1+rf'hA_1+2rfhA_1'\right)G_{3,X}
-2rf\{2rff''h-f'h(rf'-2f)
\right.\notag\\
&&\hspace{2.4cm}\left.
+fh'(rf'+2f)\}G_{4,X}
+2h\{rf^2A_1(rf'+2f)(h'A_1+2hA_1')-rA_0(rf'-2f)(2fA_0'-f'A_0)
\right.\notag\\
&&\hspace{2.4cm}\left.
+4f^2h(rf'+f)A_1^2\}G_{4,XX}
+fh\{r(2ff''-f'^2)hA_1+ff'(2hA_1+3rh'A_1+2rhA_1')\}G_{5,X}
\right.\notag\\
&&\hspace{2.4cm}\left.
-h^2A_1\{rf'^2A_0^2+rf^2f'h'A_1^2+(4f^2A_0-2rff'A_0)A_0'+2f^2f'(rhA_1A_1'-2X)\}G_{5,XX}
\right.\notag\\
&&\hspace{2.4cm}\left.
-4f^2h^2A_0'^2\left(G_{6,X}-hA_1^2G_{6,XX}+2rA_1g_{5,X}\right)
\right]\,.\label{defC13}
\eea
%

\section{
Summary of the main results}
\label{appenD}

In order to make it convenient for the readers to look for the main results of calculations in this paper, we summarize them into two tables below. The word ``stable'' means that 
the corresponding BH solution satisfies the conditions 
for the absence of ghosts and Laplacian instabilities 
associated with odd-parity perturbations. 
We caution that this does not necessarily guarantee 
the complete stability of BHs.
In case of the cubic couplings $G_3(X)$, as discussed 
in Sec.~\ref{GRG3}, the quantities 
$q_1,q_2,c_{r1},c_{r2},c_{\Omega \pm}^2$ are simply the same as those in GR, so we do not list this case in tables.

\renewcommand{\arraystretch}{3.5}
\begin{table}[htbp]
  
  \centering  
  \fontsize{9}{6}\selectfont  
  \caption{Summary of the stabilities of exact BH 
  solutions discussed in Sec.~\ref{sec4}.} 
  \label{tab1}  
    \begin{tabular}{|c|c|c|c|c|c|c|c|c|c|c|}  
    \hline  
    &  
    \multicolumn{2}{c|}{$G_4(X)$}&\multicolumn{1}{c|}
    {$G_5(X)$}&\multicolumn{2}{c|}{$G_6(X)$}
    &\multicolumn{1}{c|}{$g_4(X)$}
    &\multicolumn{2}{c|}{$g_5(X)$}
    \cr\hline   
    Forms of couplings&(\ref{func1exG4})&(\ref{func2exG4})&(\ref{funcexG5})&(\ref{funcexG6})&Arbitrary
    &(\ref{funcexg4})&(\ref{g5form})&Arbitrary
    \cr\hline   
    Branches&$A_1\neq 0$&$A_1=0$&$A_1 \neq 0, A_1=0$&$A_1=0$&$A_0'=0$&$A_1 \neq 0$&$A_1 \neq 0$&$A_0'=0$
    \cr \hline   
    BH solutions&(\ref{sol1exG4})&(\ref{sol2exG4})&(\ref{solexG5})&(\ref{sol1exG6})&(\ref{sol2exG6})&(\ref{sol1exg4})
    &(\ref{RNso})&(\ref{sol2exG6})
    \cr\hline  
    Stability conditions&$q_1q_2c_{\Omega+}^2<0$, unstable&stable&stable&stable&(\ref{G6Xre}), (\ref{G6Xre2})&(\ref{g4Xc})&stable&(\ref{g5Cond0})-(\ref{g5Cond2})
    \cr\hline  
    \end{tabular}  
\end{table}  

\begin{table}[htbp]
  
  \centering  
  \fontsize{9}{6}\selectfont  
  \caption{Summary of the stabilities of numerical BH 
  solutions discussed from Sec.~\ref{sec5} to \ref{sec6}.}  
  \label{tab2}  
    \begin{tabular}{|c|c|c|c|c|c|c|c|}  
    \hline  
    &\multicolumn{3}{c|}
    {$G_4(X)=M_{\rm pl}^2/2+\beta_4M_{\rm pl}^2 
    \left(X/M_{\rm pl}^2 \right)^n$}&\multicolumn{4}{c|}{$G_6(X)=(\beta_6/M_{\rm pl}^2) \left(X/M_{\rm pl}^2 \right)^n$}\cr\hline    
     Branches&$A_1\neq 0$, $r\sim r_h$&$A_1=0$, $r\sim r_h$&$A_1=0$, $r\gg r_h$&\multicolumn{3}{c|}{$A_1=0$, $r\sim r_h$}&$A_1=0$, $r\gg r_h$\cr\hline  
     Subcases&\multicolumn{3}{c|}{None}&$~n=0~$&$~n=1~$&$~n\geq 2~$&None\cr\hline 
   BH solutions&(\ref{fhA0hor})-(\ref{A1exrh})&(\ref{fhA0hor}), (\ref{f1h1})&(\ref{fla})-(\ref{A0la})&(\ref{fhA0hor}), (\ref{a2def})&\multicolumn{2}{c|}{(\ref{fhA0hor}), (\ref{f1h1co})}&(\ref{f0G6})-(\ref{A0P})\cr\hline  
    Stability conditions&$q_1q_2c_{\Omega+}^2<0$, unstable&stable&(\ref{noghostG4l}), (\ref{nolapG4l})&(\ref{be6con})
&(\ref{be6co})&stable&stable\cr\hline  
    \end{tabular}  
\end{table}  


%
\section*{Acknowledgements}

We thank Lavinia Heisenberg for useful discussions. 
RK is supported by the Grant-in-Aid for Young Scientists B 
of the JSPS No.\,17K14297. 
MM is supported by FCT-Portugal through 
Grant No.\ SFRH/BPD/88299/2012. 
ST is supported by the Grant-in-Aid for Scientific Research 
Fund of the JSPS No.~16K05359 and 
MEXT KAKENHI Grant-in-Aid for 
Scientific Research on Innovative Areas ``Cosmic Acceleration'' (No.\,15H05890).
YZ is supported by the NSFC grant No.\ 11605228, 11673025 and 11720101004.


\end{document}